\documentclass[11pt]{article}
\usepackage{jheppubmod}

\usepackage{amssymb, amsmath, amsopn, amsthm}

\usepackage{epsfig}
\usepackage[punctsep]{collref}
\collectsep[]{;}

\usepackage{graphicx}

\title{ CMB From CFT}

\author[1]{Ishan Mata,}
\author[2,3,4]{Suvrat Raju}
\author[1]{ and Sandip P.  Trivedi}

\affiliation[1]{Tata Institute Of  Fundamental Research,1 Homi Bhabha
  Road, Colaba, Mumbai 400 005, India.}
\affiliation[2]{International Centre for Theoretical Sciences, TIFR, IISc Campus,
  Bangalore 560012, India.}
\affiliation[3]{Harish-Chandra Research Institute, Jhunsi,
Allahabad 211019, India.}
\affiliation[4]{School of Natural Sciences, Institute for Advanced Study,
  Princeton, NJ 08540, USA.}

\emailAdd{ishanmata@gmail.com}
\emailAdd{suvrat@icts.res.in}
\emailAdd{trivedi.sp@gmail.com}

\abstract{ During inflation, spacetime is approximately  described by
  de Sitter space which is 
conformally invariant with the symmetry group $SO(1,4)$. This symmetry
can significantly constrain the quantum perturbations  which arise in the inflationary epoch.
We consider a general situation 
of single field inflation and  show that the three point function involving
 two scalar modes and one tensor mode is uniquely determined, up to small corrections,  
by the conformal symmetries. Special conformal transformations play an
important role in our analysis. Our result applies only to models where the inflaton  sector also approximately preserves 
the full  conformal group  and shows that this three point function is a good way to test if
 special conformal invariance was preserved during inflation.}

\preprint{\parbox{3cm}{TIFR/TH/12-38 \\ HRI/ST/1210 \\  ICTS/2012/11}}

\def\be{\begin{equation}}

\def\ee{\end{equation}}

\def\bea{\begin{eqnarray}}

\def\eea{\end{eqnarray}}

\def\ddk{\frac{\partial}{\partial k_1}}
\def\ddkk{{\frac{\partial}{\partial k_2}}}
\def\ddkkk{{\frac{\partial}{\partial k_3}}}
\def\vk{{\vect{k}_1}}
\def\vkk{{\vect{k}_2}}
\def\vkkk{{\vect{k}_3}}

\newcommand{\norm}[1]{#1}

\def\eepsilon{b}

\newcommand{\vect}[1]{{\boldsymbol{#1}}}

\def\la{\lambda}
\def\lad{\hat{\lambda}}
\def\lbd{\hat{\bar{\lambda}}}

\def\lb{\bar{\lambda}}

\def\mb{\bar{\mu}}

\def\ep{\epsilon}

\def\Or[#1]{{\text{O}}\left({#1}\right)}
\def\dotl[#1,#2]{\left\langle #1,\, #2 \right\rangle}
\def\dotlb[#1,#2]{\left\langle #1,\, #2 \right\rangle}
\def\dotlm[#1,#2]{\left[ #1,\, #2 \right]}
\def\dotp[#1,#2]{(\vect{#1} \cdot\vect{#2})}
\keywords{inflation, correlation functions,  de Sitter space, conformal field theory}
\begin{document}

\maketitle

\section{Introduction}
Inflation  states that 
  our   Universe  underwent a period of  exponentially rapid 
 expansion in its early history. This idea solves the flatness and 
horizon problems in cosmology.  
  What is particularly attractive  is that the same exponential expansion also results in 
small quantum perturbations being produced which
account for  the observed anisotropies of the microwave background and also provide the seed perturbations for the growth of large scale structure in the Universe.

The exponentially expanding Universe during inflation is well described by the metric of de Sitter space, up to 
small corrections.  
It is well known that de Sitter space is a maximally  symmetric spacetime. In four dimensions the group of 
isometries of de Sitter space is $SO(1,4)$ --- the Lorentz group in $4+1$ dimensional flat spacetime. 
This large group of symmetries has ten generators, which include translations and rotations along the 
three space directions, 
 scale transformations, and the three generators  of  special conformal transformations. 
We will refer to it as the conformal group below. 

So far, the experimental tests of inflation,  coming for example  from the study of the CMB, 
have shown that the perturbations can be  well approximated as being    Gaussian. 
The good news is that future experiments,  with improved sensitivity, will be able to probe and possibly 
 detect evidence for    non-Gaussianity in these perturbations. 
For example, it is hoped that the Planck experiment will be able to provide significant constraints of this sort
 quite soon. 

A Gaussian distribution is completely determined  by its two point correlation  function.
 Any  non-Gaussianity in the perturbations can  therefore be characterized  by the  three point or 
 higher point correlations.
 Considerable attention  has been paid in the recent literature to the three-point function, called 
the bispectrum; there is also a growing body of literature on the four point function, called the trispectrum. 
We refer
the reader to
\cite{Komatsu:2010hc,Komatsu:2009kd,Bartolo:2004if,Komatsu:2002db} 
 for a review of these developments and to  \cite{Weinberg:2008zzc}
 for background material. 

There are two kinds of perturbations of the metric  that are relevant  for inflation: these transform as scalars
and spin-2 representations of the rotation group, and are called scalar and tensor perturbations respectively.
In addition each perturbation is characterized by a value for the  spatial three-momentum. 
It is easy to see that the momentum dependence of the two-point function of the perturbations is simple and 
 is fixed, up to small corrections,  by the approximate scale invariance of de Sitter space.  
 One the other hand, it is well known that   the momentum dependence of the three point functions can be much more complicated. For example, various different shapes which characterize this momentum dependence have been 
obtained for the three point scalar correlation function in different
models of inflation. (See \cite{Komatsu:2010hc,
  Chen:2006nt} and references there.)

The symmetries of de Sitter space need not be shared by the scalar sector in general. 
This happens for  example in DBI inflation
\cite{Silverstein:2003hf,Alishahiha:2004eh}
where the non-canonical kinetic energy term for the inflaton results
in a  speed of sound $c_s\ne 1$.\footnote{Another example where the scalar sector violates
special conformal symmetries is ghost inflation
\cite{ArkaniHamed:2003uz}.} As a result, while scaling symmetry is preserved,
 the inflaton sector breaks special conformal invariance  badly. 
Here we will assume that the full conformal group is approximately preserved by the inflationary dynamics, 
including both gravity and the inflaton field, and examine the resulting constraints imposed on three point 
functions. 
 
In particular, we will focus on the three point function involving two scalar perturbations $\zeta(\vect{k})$
 and one tensor perturbation $\gamma_{ij}(\vect{k})$, with
 polarization $e^{s, ij}$, denoted by,
\be
\label{defcorr}
\langle\zeta(\vect{k}_1)\zeta(\vect{k}_2) \gamma_{ij}(\vect{k}_3)\rangle e^{s,ij}.
\ee
We will   show that this correlator is completely fixed  by symmetry considerations.\footnote{A complete complete definition
of the perturbations etc. is given in section \ref{basicsetup}.}
Its  overall  normalization  is determined
 in terms of the two point functions of the scalar and tensor perturbations,
and its momentum dependence is  determined by the $SO(1,4)$ symmetry group. 
It turns out that the special conformal transformations play an   especially important role in our
 analysis.  They give rise to differential equations for  the correlation function whose solution
 is essentially unique leading to the conclusion above. 
In the absence of special conformal invariance in the full theory, including the inflaton sector, our results for the correlator are not valid. 

Our analysis applies to models with  only one scalar field during inflation. It also assumes that the 
initial state was the Bunch-Davies vacuum.\footnote{These boundary conditions are restated in a way more convenient for our analysis in section \ref{symandcons}.}
Beyond that, our analysis only relies on the conformal group  and is essentially model independent.
In particular, our results also  apply to models where higher derivative corrections are important and 
gravity or the scalar field   is not well described by the two-derivative approximation. In the context of string theory, such a situation would arise  if the Hubble scale $H$
 during inflation was of order the string scale $M_{st}$. Present bounds on $H$ coming, for example, from the 
 absence of any observed effects due to tensor perturbations tell us that $H <\sim 10^{16} \text{Gev} < M_{Pl}$.
So, for example,  the  higher derivative corrections would be important if   $H$ and 
$M_{\text{st}}$ are both  comparable and of order the Grand unification scale $ M_{\text{GUT}}\sim 10^{16} \text{Gev}$.
Since very little is understood about string theory in time dependent backgrounds  the 
resulting correlation functions in such a situation    cannot be calculated directly 
from our present knowledge of the theory. 
However symmetry considerations   still hold and our result for the correlation function \eqref{defcorr}
 is valid for   such  a situation as well.

The generality of our  result makes the correlator given in  \eqref{defcorr}  a good test, 
in a model independent manner,  of the full symmetry group during inflation. 
The two-point scalar correlator, which has now been measured, is consistent with approximate scale 
invariance but this leaves open the possibility that the special conformal symmetries of 
de Sitter space are not preserved by the scalar sector. In fact, as was mentioned above, it is easy enough to construct models of inflation 
where this does happen and also straightforward to see that this possibility is allowed in terms of an effective field theory analysis
\cite{Cheung:2007st}. 
The correlator discussed here, if observationally measured,   can  conclusively settle 
whether the special conformal symmetries were approximately preserved  during inflation. 

Unfortunately, experimental tests of this three point correlator 
 are  still some way away  since its  magnitude is small. 
Even the detection of the 
two point function for the tensor mode has not been made so far  and would be a great discovery in
 itself.  The small value that the three point scalar correlator has in conventional slow-roll inflation
 can be enhanced  in models like DBI inflation which involve  the breaking of  
 special conformal symmetries. 
However, with the special conformal symmetries intact  our analysis fixes the the overall
normalization of the correlation function with two scalars and one tensor,  as was mentioned above, and  
rules out the possibility of any such  enhancement.

Therefore, we present   the result of our analysis here not with any immediate experimental contact in mind,
but rather  with a view to the future when hopefully such contact will become possible and such model independent tests of inflation might play a useful role in sharping our understanding of the early Universe. 

A second motivation for our work comes from the study of conformal field theory.
The symmetry group mentioned above, $SO(4,1)$, is exactly the same as the symmetry group 
of a $3$ dimensional Euclidean conformal field theory (CFT). 
This is in fact why we referred to this symmetry group as the conformal group when we first introduced
 it above.  The problem of studying the constraints imposed by this symmetry group  on 
the correlation functions of the scalar and tensor perturbations in de Sitter space
maps in a direct way to the question of studying the constraints imposed in a $3$ dimensional 
conformal field theory on correlation functions involving  a nearly marginal  scalar operator
and the stress energy tensor. Thus our analysis is also of interest in  the study of 3 dimensional CFTs: a subject
which has also been of some considerable interest recently.\footnote{For some discussion of 
three-point functions in 3 dimensional CFTs see 
\cite{Osborn:1993cr,Giombi:2011rz,Maldacena:2012sf,Maldacena:2011jn}.}

The three point correlation function for two scalar operators  and the  stress tensor
is already well known  in the CFT literature \cite{Osborn:1993cr}.
However this result is in position space, while
for cosmology one is interested in the answer in momentum space. It is
not easy to directly Fourier transform the position space
result. Moreover, the position space answer has divergences where the
operators come together. It is rather subtle to regulate these
divergences --- which is necessary to define the Fourier transform ---
while preserving conformal invariance.  A closely related issue is
that of contact terms, which can also arise in position space. These
were not determined in \cite{Osborn:1993cr} but are important for the momentum dependence of the correlator.  
As our analysis shows, working directly in momentum space, the symmetry considerations are powerful enough
 to fix these ambiguities for the correlator and  determine  a unique answer.

Finally, a third motivation comes from attempts to study de Sitter space and its possible dual description in 
terms of a CFT
\cite{Strominger:2001pn,Witten:2001kn,Maldacena:2002vr,Anninos:2011ui}.
It is unclear at this point  whether  a precise correspondence of this type is possible. 
However, symmetry properties for correlators can be related between the gravity description and the CFT,
 as mentioned above. These are analogous to and in fact  follow after  analytic continuation from
the correspondence between correlators in the AdS/CFT case.  Since, as our results help show,
 symmetry properties can significantly constrain at least some of the correlators,
 the correspondence in this limited sense is still 
of some practical benefit. 

Before going further we must mention the seminal  papers  of 
Maldacena \cite{Maldacena:2002vr} and more recently  Maldacena and
Pimentel \cite{Maldacena:2011nz}.
These papers lay out  the essential ideas on which 
our analysis is based. The precise nature of the map between the gravity theory and the CFT using the 
wave function of the Universe was first discussed in
\cite{Maldacena:2002vr}.
 And the importance of special 
conformal transformations was discussed in \cite{Maldacena:2011nz}
 where  it was also shown that these symmetries 
significantly constrain the three point
function of  tensor perturbations. Our analysis  is  a modest extension of this approach 
for a  correlator involving scalar perturbations as well.

Other relevant  works which explore similar ideas are 
\cite{Antoniadis:2011ib,Bzowski:2011ab,McFadden:2011kk,McFadden:2010vh,Larsen:2003pf,Larsen:2002et,
Cheung:2007st,Weinberg:2008hq}.
Two recent papers \cite{Schalm:2012pi,Bzowski:2012ih} appeared while this paper  was being prepared for publication and contain related material.  

This paper is organized as follows. In \S2 we discuss the basic ideas behind the analysis and background 
material. In \S3 we  set up the equations which arise due to conformal invariance. 
In \S4 we discuss  a solution to these equations and prove  that it is unique.
 Our final results are presented in \S5. We end with conclusions in \S6. 
Three  Appendices contain important supplementary material follow. 
A reader who is not interested in the details of the calculations can
read the introduction, and then turn directly  to \S5 with the final results, which can be read in a self contained way together with Appendix A,
and then end with   the conclusions.

\section{Basic Set-Up}
\label{basicsetup}

We consider a theory of gravity coupled to a scalar field, the inflaton, with action
\be
\label{action1}
S=\int d^4x\sqrt{-g}{1\over 16 \pi G} [R-{1\over 2} (\nabla \phi)^2   -V(\phi) + \cdots].
\ee
The ellipses  stand for  higher derivative corrections involving, in general,
 both gravity and the inflaton. Such corrections could be important, for example, 
 if the Hubble scale during inflation is of order the string scale. 
Note that in \eqref{action1} we are using conventions where the inflaton is dimensionless. 
Also below we will choose
conventions where the Planck scale   
\be
\label{defmp}
M_{Pl}^2 \equiv 8 \pi G = 1.
\ee
It is well known that during inflation the Universe is approximately
described by de Sitter space
\begin{eqnarray}
&&ds^2 =-  dt^2 + a^{2}(t) \sum_{i=1}^3 dx_i dx^i, \label{ds} \\
&&a^2  =  e^{2 Ht}, \label{scaleds}
\end{eqnarray}
 and hence undergoes exponential expansion. In  \eqref{scaleds}, $H$ is the Hubble scale which is a constant in 
de Sitter space.  The inflationary epoch is described by de Sitter space with small corrections.
 These arise   because   of the  slow variation of the Hubble scale which can  
 be parametrized
in terms of the  two parameters
\be
\label{srp}
\epsilon=-{\dot{H}\over H^2}, \delta = {\ddot{H}\over 2 H \dot{H}},
\ee
where dot denotes derivative with respect to $t$. 
During   inflation both these parameters are small and 
  meet the slow roll conditions
\be
\label{src}
\epsilon, \delta \ll 1.
\ee

When the two-derivative approximation is good and the action can be approximated by the terms given in 
\eqref{action1},
 $H$ is given in terms of $V$ by 
\be
\label{relhv}
H=\sqrt{V \over 3 M_{Pl}^2},
\ee
and the slow roll parameters can be expressed in terms of  of $V$ by 
\begin{align}
&\epsilon  =  {1\over 2} {M_{pl}^2 (V')^2 \over V^2}  \label{relsla}, \\
&\delta  =   -M_{pl}^2 {V''\over V} +\epsilon, \label{relslb}
\end{align}
where  prime denotes derivatives with respect to the scalar field.\footnote{The slow-roll parameter $\eta$ which is more conventionally used is given by $\eta= M_{Pl}^2 {V''\over V}$.}
Also in the two-derivative theory we have 
\be
\label{relsla2}
\epsilon={1\over 2} {\dot{\phi}^2\over H^2}.
\ee   
When the two-derivative approximation  is not valid  $\epsilon$ defined in \eqref{srp}
 and $\dot{\phi}$ will not be related by \eqref{relsla2}
in general. The slow-roll approximation then  requires that besides \eqref{src} being valid,
  \be
\label{srscalar}
{\dot{\phi}\over H} \ll 1.
\ee 

de Sitter space is well known to be conformally invariant. For example it is easy to see that the scale 
 transformation
\be
\label{scinv}
x^i \rightarrow \lambda x^i, t \rightarrow t - {1\over H } \log(\lambda),
\ee
leaves the metric \eqref{ds} invariant. 
More generally the full isometry group of de Sitter space is $SO(1,4)$. It consists of the usual three translations and
 rotations in the $x^i$ coordinates, the scale transformation, \eqref{src}, and in addition three special conformal 
transformations. Infinitesimal special conformal   transformations are of the form
\begin{align}
&x^i   \rightarrow   x^i -2 (\eepsilon_j x^j) x^i + \eepsilon^i (\sum_j (x^j)^2 - e^{-2Ht}) \label{spcon}, \\
&t  \rightarrow   t + 2 \eepsilon_j x^j \label{tspcon}. 
\end{align}
Here $b^i, i =1, \ldots 3$ are infinitesimal parameters.  
As mentioned above  de Sitter space is modified during inflation 
 due to the time  varying  Hubble scale. While translations and
 rotations in the $x^i$ directions are of course unbroken, this
 modification results in  the breaking of the scaling and special conformal symmetries.
However, as long as the  slow roll parameters $\epsilon, \delta,$ are
small this breaking is small and the resulting inflationary spacetime
is still approximately conformally invariant. 

The inflaton sector need not preserve the full conformal group breaking the $SO(1,4)$ symmetry of
 de Sitter space badly and only preserving translations, rotations and scale transformations, as was 
mentioned in the introduction. Additional parameters enter in such a model 
 which parameterize this breaking. 
For example, the speed of sound, $c_s$, is one such parameter. When $c_s\ne 1$ the special conformal symmetries are broken. See \cite{Cheung:2007st} for a more general parametrization of such effects. 
In what follows we will assume that the scalar sector also approximately preserves the full symmetry group
of de Sitter space.

\subsection{The  Perturbations}
\label{theperturbations}

The inflationary space-time  is a solution for the  system consisting of gravity and a scalar field. The rotational 
invariance in the $x^i$ directions can be used to characterize perturbations about this solution. 
There are two kinds of perturbations which can arise, scalar  and tensor perturbations. The scalar perturbations have   spin zero 
and the tensor perturbations have  spin 2. 

The tensor perturbations are easy to understand ---  they are gravity waves in the inflationary background. 
The scalar perturbations essentially arise due to the presence of the inflaton field. 
Depending on the gauge chosen they can be thought of as perturbations in the inflaton, or in the spatial curvature or in 
a combination of both of these modes. 

\subsubsection{Gauge 1}
\label{gauge1}
For example,  we can choose a gauge where the perturbations in the inflaton vanish,
\be
\label{perti}
\delta \phi =0.
\ee
Starting with the form of the metric  used in the ADM formalism
\be
\label{admmetric}
ds^2=-N^2 dt^2 + h_{ij} (dx^i + N^i dt) (dx^j+ N^j dt),
\ee
the additional coordinate  reparameterization can be fixed by choosing a gauge where 
\be
\label{gfversion}
 h_{ij}=a^2 [(1+ 2\zeta) \delta_{ij} + \gamma_{ij}],
\ee
where $\gamma_{ij}$ is transverse and traceless,
\be
\label{tt}
\partial_i \gamma_{ij} =\gamma_{ii}=0,
\ee
as discussed in \cite{Maldacena:2002vr}. The tensor perturbations are given by $\gamma_{ij}$.
And the scalar perturbations are  given by $\zeta$ and correspond to fluctuations in the spatial curvature 
along the spatial directions. 

\subsubsection{Gauge 2}
\label{gauge2}
Alternatively, for the scalar perturbations, we can choose to set 
$\zeta$ instead of $\delta \phi$ to vanish. The perturbations are now given by 
fluctuations in the inflaton, $\delta \phi$.
This second gauge is obtained by starting with the coordinates in which the perturbations take the form given in the previous 
paragraph, $\zeta, \gamma_{ij}$ and carrying out a time reparameterization 
\be
\label{tre}
t\rightarrow t+ {\zeta  \over H}.
\ee
It is easy to see that this sets $\zeta$ to vanish. The tensor perturbation $\gamma_{ij}$ is unchanged by this coordinate transformation.
If the background value of the inflaton in the inflationary solution is 
\be
\label{backinfl}
\phi=\bar{\phi}(t),
\ee
the resulting value for the perturbation $\delta \phi$ this gives rise to is 
\be
\label{infper}
\delta \phi= -{\dot{\bar{\phi}}  \zeta \over H}. 
\ee
When the two derivative approximation is good  we can using \eqref{relsla2}  express this relation as 
\be
\label{infper2}
\delta \phi=-\sqrt{2 \epsilon} \zeta.
\ee
We will find it useful to consider both gauges in our discussion below. 
 As we will discuss further in subsection \ref{symandcons} for our purposes it will be most
 convenient to first work in gauge 2, where the scalar perturbation is given by $\delta \phi$ and 
then transform to gauge 1,
where the perturbation is given by $\zeta$, around the time when the mode crosses the horizon.
This might seem conceptually confusing at first but has the advantage of allowing us to incorporate both
 the leading effects of the slow-roll parameters in a straightforward manner and of eventually going over to the description in terms of $\zeta$ which is 
the variable   that it is defined for all time and also becomes  constant once the mode exits the 
horizon. 

Let us also make one more comment here. 
The relation \eqref{infper} has
corrections involving higher powers of the perturbation, $\delta \phi$. 
For 
the scalar three-point function in  conventional  slow-roll models, as studied in \cite{Maldacena:2002vr},
 the first corrections to \eqref{infper} 
need to be kept since the leading answer is suppressed by an additional
power of $\sqrt{\epsilon}$.  But these corrections  can be ignored for the correlator \eqref{defcorr}.

\subsection{The Wave Function}
The time dependence during the inflationary epoch gives rise to scalar and tensor perturbations. 
Our main interest in this paper   is to ask about the constraints that  approximate conformal invariance imposes on the 
correlation functions of these  perturbations.
In particular we will be interested in these  correlation functions at late enough times
 when the modes  have crossed the horizon,
and  their wavelength, $\lambda$,  has become much bigger than the Hubble scale, $\lambda \gg H^{-1}$. 

At such late times the correlations functions acquire a time independent limiting form.
The physical reason for this is well understood. Once the wavelength of a mode gets much longer than the Hubble scale
the evolution of the mode gets dominated by Hubble friction and as a result it comes to rest. 

In our discussion it will be  useful to think in terms of a  wavefunction which describes the state of the system at late times.
The wavefunction tells us the amplitude to observe a particular perturbation and clearly 
encodes all information about   the correlation functions.
Since the correlation functions become time independent at late times  the wave function also becomes time independent
in this limit.\footnote{More accurately, this happens after suitable infra-red divergences are subtracted.
 Physical answers do 
not depend on the choice of subtraction procedure.}

The wave function will be a convenient description for our analysis since we are interested in the constraints imposed by 
symmetries and these can be conveniently translated to invariances of the wavefunction as we will see shortly. In turn this will 
allow us to map the constraints imposed by symmetries to  an analysis of constraints imposed on correlators in a 3 dimensional 
Euclidean conformal field theory.  More generally, thinking in terms of the wave function  also allows
 us to exploit the analogy with calculations in AdS space for our purpose. 

The perturbations produced  during inflation are known to be Gaussian with small corrections. This allows the 
late time wave function to be written as a power series expansion of the form
\be
\label{wf1}
\begin{split}
\psi[\chi(\vect{x})]  = & \exp\bigl(-{1\over 2} \int d^3 x d^3 y \chi (\vect{x}) \chi(\vect{y}) \langle\hat{O}(\vect{x}) \hat{O}(\vect{y})\rangle  \\
&+ {1\over 6} \int d^3 x d^3 y d^3 z\, \chi(\vect{x}) \chi(\vect{y})\chi(\vect{z}) 
\langle\hat{O}(\vect{x}) \hat{O}(\vect{y}) \hat{O}(\vect{z})\rangle + \cdots \bigr). 
\end{split}
\ee
Here $\chi$ stands for a generic perturbation which could be a scalar or tensor perturbation. The ellipses stand for higher order terms involving more powers of $\phi$. The coefficients $\langle\hat{O}(\vect{x}) \hat{O}(\vect{y})\rangle, \langle\hat{O}(\vect{x}) \hat{O}(\vect{y}) \hat{O}(\vect{z})\rangle$ etc. 
are for now just functions which determine
the correlators.

The expression above is schematic. In the case at hand there are two kinds of perturbations, scalar and tensor. 
Working in the gauge described in subsection  \ref{gauge2} these are 
$\delta \phi, \gamma_{ij}$. With a suitable choice of normalization
the wave function will then take the form
\be
\label{wf2}
\begin{split}
\psi[\delta \phi, \gamma_{ij}] & =  \exp\bigl[{M_{pl}^2 \over H^2} \bigl(-{1\over 2} \int d^3 x d^3 y \delta \phi(\vect{x}) \delta \phi(\vect{y}) \langle O(\vect{x}) O(\vect{y})\rangle  \\
&- {1\over 2} \int d^3 x d^ 3 y \gamma_{ij} (\vect{x}) \gamma_{kl}(\vect{y}) \langle T^{ij}(\vect{x}) T^{kl}(\vect{y})\rangle  \\  
& -{1\over 4} \int d^3 x d^3 y d^3 z \delta \phi (\vect{x}) \delta \phi(\vect{y})
\gamma_{ij}(\vect{z}) \langle O(\vect{x}) O(\vect{y}) T^{ij}(\vect{z})\rangle + \cdots \bigr)
\bigr].
\end{split}
\ee

The ellipses stand for additional terms of various kinds involving three powers of the perturbations with 
appropriate coefficient functions  and then higher order terms. 
  
Note, in our notation every additional power of the scalar perturbation is accompanied by an additional 
factor of 
$O(\vect{x})$ in the coefficient functions and  every additional power of the tensor perturbation is 
accompanied by an additional
 factor of $T_{ij}(\vect{x})$. We will soon see that the coefficient functions transform under the  symmetries in the same  way as correlation functions  involving a scalar operator and the stress energy tensor in a $3$ dimensional Euclidean conformal field theory.   

In this paper our interest will be on the last term in the RHS of \eqref{wf2}.
 Together with the two point functions,  this term determines the three point correlator of interest to us.

\subsection{Symmetries and Their Consequences}
\label{symandcons}

We have seen that the wave function at late times is a functional of the late time values of the  
perturbations.   
Schematically we can write 
\be
\label{wf3}
\psi[\chi(\vect{x})]=\int^{\chi(\vect{x})} D \chi e^{i S},
\ee
where $\chi$ again stands for the value a generic perturbation takes at late time and 
the action for any configuration is denoted by $S$. 
We would now like to derive constraints imposed by symmetries on this wavefunction. 

Before doing so it is worth considering the boundary conditions in the path integral in more detail. 
We will consider inflation with the  standard Bunch-Davies boundary conditions in the far past, when the modes of interest
had a wavelength much shorter than the Hubble scale. At these early times the short wavelengths of  the modes
makes them  insensitive to the geometry of de Sitter space and   they essentially propagate as if  in Minkowski spacetime.
The Bunch Davies vacuum corresponds to taking the modes to be in the Minkowski vacuum at early enough time.

An elegant way to impose this boundary condition in the path integral above, as discussed in  
\cite{Maldacena:2002vr}, is as follows.
Consider de Sitter space in conformal  coordinates,
\be
\label{poincoord}
ds^2={1\over \eta^2} (-d\eta^2 + (dx_i)^2),
\ee
with the far past being $\eta \rightarrow -\infty$, and late time being $\eta \rightarrow 0$. 
Continue $\eta$ so that it acquires a small imaginary part $\eta \rightarrow \eta (1- i \epsilon), \epsilon > 0$.
Then the Bunch Davies boundary condition is correctly imposed if the path integral is done over configurations 
which vanish at early times when $\eta \rightarrow -\infty(1- i \epsilon)$. 
Note that in general the resulting path integral is over complex field configurations. 

As an  example, consider   a free field $\phi$ satisfying  the equation
\be
\label{eqff}
\nabla^2 \phi=0.
\ee
A mode with momentum $\vect{k}$ is of the form, $\phi=f_{\vect{k}}(\eta) e^{i\vect{k} \cdot \vect{x}}$, where
\be
\label{valfeta}
f_{\vect{k}} = c_1 (1-i k \eta) e^{i k \eta} + c_2 (1+ i k \eta) e^{-ik\eta},
\ee
and $k \equiv |\vec{k}|$.
Requiring that the solution vanish when $\eta \rightarrow -\infty(1-i\epsilon)$, sets $c_2=0$ and requiring $f_{\vect{k}}$ equals 
the boundary  value,
$f_{\vect{k}}=f_{\vect{k}}^0$ at the 
late time $\eta=\eta_c$, gives
\be
\label{eqff2}
f_{\vect{k}}=f^0_{\vect{k}}{(1-i k \eta) e^{i k \eta} \over (1-i k \eta_c) e^{i k \eta_c}}.
\ee
Since $f_{\vect{k}}\ne f_{-\vect{k}}^*$ the resulting field configuration is complex. 

We are now ready to return to our discussion of the  constraints imposed by symmetries on the wave function. 
What is important for this purpose, as far as the boundary conditions in the far past are concerned, 
 is that the field configurations we sum over in the path integral vanish in the far past. 

 Consider in fact a general 
situation where we have a wave function of the form \eqref{wf3}
 for a general set of fields $\chi$, with some boundary condition in the far past.
  Now if the system has a symmetry which keeps the action and the measure invariant and which also preserves the
 boundary conditions in the far past and if under the symmetry 
the boundary value of the field $\chi$ transforms as follows
\be
\label{symtr}
\chi(\vect{x}) \rightarrow \chi'(\vect{x}),
\ee
then it follows from the definition of the wave function \eqref{wf3} that  $\psi[\chi]$
satisfies the condition
\be
\label{condsymwf}
\psi[\chi(\vect{x})]=\psi[\chi'(\vect{x})],
\ee
and is invariant under the symmetry.

For the case at hand where we work with de Sitter space, the symmetry group is the conformal group $SO(1,4)$ 
of isometries discussed above. Being isometries,  the action and measure are invariant under it on account of 
reparameterization invariance. 
The boundary condition in the far past corresponding to the Bunch Davies vacuum  is that the fields vanish.
This is indeed preserved
by the conformal transformations since the field transform homogeneously under these symmetries. 
For tensor perturbations this is all we need to use the general argument above. It follows that the wave function must be invariant under a change of the 
boundary values of the tensor perturbations which arise due to conformal transformations. 
As we will see shortly this implies that the coefficient functions, which we have suggestively 
denoted  as $\langle T_{ij} T_{kl}\rangle$ etc.,  behave exactly like the correlations functions of the stress energy tensor of a 
three  dimensional conformal field theory under conformal transformations. 
It is true, as  we discussed above, that conformal invariance is broken slightly during inflation but this leads to 
only subleading corrections in the tensor mode correlations. 

For the scalar mode the situation  is a little more complicated. In pure de Sitter space, without the  inflaton, 
the  scalar perturbation in the metric $\zeta$, \eqref{admmetric},  is pure gauge.  In the presence of the inflaton there is a genuine
 scalar perturbation. However as \eqref{infper}, \eqref{infper2}
  which relates the perturbations in the two gauges discussed 
in section \ref{theperturbations}  shows, the slow roll parameter $\epsilon$ which is non-zero due to the
 breaking of conformal invariance is  then    involved in the  
definition of the scalar perturbation itself. This can make it confusing to apply the consequences of the small breaking of 
conformal invariance to the scalar sector. 

 The simplest way to proceed is to work in the second gauge  discussed in subsection \ref{gauge2},  where $\zeta=0$. 
The scalar perturbation is then 
just the fluctuation in the scalar field. To leading order in the slow-roll parameters these fluctuations can be calculated in de Sitter space and the time evolution of the inflaton can be neglected for this process. As a result the full set of perturbations,
 scalar and tensor, with Bunch-Davies boundary conditions, then meet the conditions of the general argument given above and 
we learn that the wave function must be invariant under conformal transformations of the boundary values of these perturbations. 

Once the results are obtained in this gauge one can always transform to other gauges, in particular the first gauge considered in subsection \ref{gauge1}  where $\zeta$ is non-vanishing. In fact this is very convenient to do 
for purposes of following the evolution of the scalar mode  after the end of inflation. Since $\zeta$ is related 
to $\delta \phi$ by \eqref{infper} the  resulting correlation functions will depend
 on the breaking of conformal invariance  even to leading order 
 but this dependence  arises solely due to the relation \eqref{infper}  and is easily obtained. 

Before proceeding let us note that the discussion above has a direct parallel with what happens in a conformal field 
theory which is deformed by adding a perturbation
\be
\label{percft}
\delta S= \int g O,
\ee
 which breaks conformal symmetry slightly. 
Due to this breaking the trace of the stress tensor $T^i_i$ does not vanish anymore and instead satisfies the relation
\be
\label{tracerel}
T^i_i=\beta(g) O,
\ee
where $\beta(g)$ is the beta function for the coupling in \eqref{percft}. To leading order in the breaking
correlation functions for $T^i_i$ can be obtained by first calculating those of $O$ in the CFT (without any breaking)
and then transforming these to correlation functions for $T^i_i$ using \eqref{tracerel}. 

\subsection{Constraints on Coefficient Functions}
Let us now work out the constraints imposed by conformal symmetries on the coefficient functions which arise in the expansion
of the wave function \eqref{wf1}  in more detail. 
It is easy to see that the constraints of translational invariance make the coefficient functions also translationally invariant.
Under rotations in the $x^i$ directions  the  wave function will be  invariant
if $O(\vect{x})$ transforms like a scalar and $T_{ij}$ like a two-index tensor within coefficient functions. 

Next  we come to the scale transformation and special conformal transformations. 
Under the scale transformation \eqref{scinv}  
the scalar perturbation transforms by 
\be
\label{scalartr0}
\delta \phi(\vect{x},t)\rightarrow \delta \phi'(\vect{x},t)= \delta \phi({\vect{x} \over \lambda}, t+{1\over H} \log(\lambda)).
\ee 
At late times $\delta \phi$ becomes independent of $t$, as a result this equation becomes
\be
\label{scalartr}
\delta \phi(\vect{x})\rightarrow \delta \phi'(\vect{x})= \delta \phi({\vect{x} \over \lambda}).
\ee
In particular this is true for the boundary value of $\delta \phi$ as well.

As a result, suppressing the dependence on tensor modes for the moment, we learn that the wavefunction must satisfy the conditions 
\be
\label{wfstr}
\psi[\delta \phi(\vect{x})]=\psi[\delta \phi'(\vect{x})]=\psi[\delta \phi({\vect{x} \over \lambda})] .
\ee
As mentioned above every additional factor of $\delta \phi(\vect{x})$ in the expansion of the wave function involves an additional factor
of $O(\vect{x})$ in the corresponding  coefficient function  and also an integral over the spatial position of $\delta \phi(\vect{x})$.
Thus schematically speaking the wave  function will satisfy the condition \eqref{wfstr}   if 
\be
\label{wf2str}
\int d^3x \delta \phi'(\vect{x}) O(\vect{x}) = \int d^3x \delta \phi(\vect{x}) O(\vect{x}),
\ee
where more correctly we mean the coefficient functions involving $O(\vect{x})$, rather that $O(\vect{x})$ itself.
 This leads to the condition
\be
\label{condaa}
\int d^3x \lambda^3 \delta \phi(\vect{x}) O(\lambda \vect{x}) =\int d^3 x \delta \phi(\vect{x}) O(\vect{x}).
\ee
(In deriving this relation  we first change variables in the middle expression   of \eqref{wf2str} to $\vect{y}={\vect{x}\over \lambda}$ and then change $\vect{y}$ to $\vect{x}$ since
it is a dummy variable of integration.) Since \eqref{condaa}  is true for an arbitrary function $\delta \phi(\vect{x})$ we learn that  
coefficient functions  are invariant under the replacement
\be
\label{traops}
O(\vect{x}) \rightarrow \lambda^3 O(\lambda \vect{x}).
\ee
Or in infinitesimal form if  $\lambda=1+\epsilon$,
\be
\label{trbops}
O(\vect{x}) \rightarrow O(\vect{x}) + \epsilon \delta O(\vect{x}),
\ee
with 
\be
\label{trcops}
\delta O(\vect{x})=3 O(\vect{x})+ x^i\partial_i O(\vect{x}).
\ee

This is exactly the condition that would arise due to scale invariance if the coefficient functions were   the 
correlation functions in  a conformal field theory with $O(\vect{x})$ being an operator of dimension $3$. 
Note that in $3$ dimensions this  makes $O(\vect{x})$ marginal.

A similar  argument for the tensor perturbation shows that under the scaling transformation, \eqref{scinv}, 
 the boundary value of the tensor perturbation
transforms like\footnote{The reader might find this puzzling at first since the metric should transform as a tensor
under the coordinate transformation \eqref{scinv}.
In fact the metric $h_{ij}$, \eqref{admmetric}, does transform  like a tensor and goes to
 $h_{ij}(\vect{x}) \rightarrow {1\over \lambda^2} h_{ij}({\vect{x}\over \lambda})$. However $\gamma_{ij}$ is related to $h_{ij}$ 
after multiplying by an additional factor of $a^2$, \eqref{gfversion}.
Since $t$ shifts, \eqref{scinv},  the $a^2$ factor also changes
resulting in the transformation rule \eqref{ttrans}.}
\be
\label{ttrans}
\gamma_{ij}(\vect{x}) \rightarrow \gamma_{ij}'(\vect{x})=\gamma_{ij}({\vect{x}\over \lambda}).
\ee
This is entirely analogous to \eqref{scalartr} and a similar argument leads to the 
conclusion that $T_{ij}$ must behave like an operator of dimension $3$ under scaling transformations
 for the wave function to be invariant
under it. 

Finally we consider special conformal transformations. 
At late times when $e^{-Ht}\rightarrow 0$ we see from \eqref{spcon}  that the  $x^i$ coordinates  transform as 
\begin{align}
&x^i \rightarrow x^i + \delta x^i, \label{cspct} \\
&\delta x^i  =  x^2 \eepsilon^i - 2 x^i \dotp[x, \eepsilon]. \label{cspct2}
\end{align}
Henceforth we will use  notation where $\dotp[a,b] \equiv a^i b_i$ and also  raise  and lower 
indices along the spatial
directions using the flat metric $\delta_{ij}$.

The boundary value of the scalar field perturbation transforms under this as
\be
\label{scsptr}
\delta \phi(\vect{x}) \rightarrow \delta \phi'(\vect{x})=\delta \phi(x^i -\delta x^i).
\ee
Arguing as in the case of the scale transformation above we then learn that for the wave function to be invariant
coefficient functions must be invariant when 
\begin{eqnarray}
O(\vect{x}) &  \rightarrow & O(\vect{x}) + \delta O(\vect{x}), \label{ospcft} \\
\delta O(\vect{x}) & = &- 6 ( \vect{x} \cdot \vect{\eepsilon}) O(\vect{x}) + D O(\vect{x}), \label{ospcft2} \\
D & = & x^2 \dotp[\eepsilon,\partial] - 2 \dotp[\eepsilon,x] \dotp[x,\partial]. \label{defD}
\end{eqnarray}
This is exactly the transformation of an operator of dimension $3$ under special conformal transformations. 
Similarly from the transformation of the tensor mode we learn that the coefficient functions must be invariant when 
\begin{eqnarray}
T_{ij}(\vect{x}) & \rightarrow &  T_{ij}+\delta T_{ij}, \label{stressspcft} \\
\delta T_{ij} & = & -6 (\vect{x}\cdot \vect{\eepsilon}) T_{ij} + 2 \hat{M}^k_i T_{kj} + 2 \hat{M}^k_j T_{ik}- D T_{ij}, \label{stressspcft2} \\
\hat{M}^k_i & \equiv &  2(x^k \eepsilon^i - x^i \eepsilon^k). \label{hatM}
\end{eqnarray}
These agree with the  transformation rules for the stress energy tensor of a 3d CFT and also agree with 
eq.(4.9) in \cite{Maldacena:2011nz}. 

The stress energy tensor of a CFT also satisfies one additional condition --- it is conserved. This gives rise to Ward identities
that must be satisfied by correlations functions in the CFT involving the stress energy tensor. 
The same conditions  also arise for  the coefficient functions at hand here.
The wave function must be reparameterization invariant with respect to general coordinate transformations, 
\be
\label{coortr}
x^i \rightarrow x^i + v^i,
\ee
under which the metric and scalar perturbations transform as
\begin{eqnarray}
\gamma_{ij} & \rightarrow &  \gamma_{ij} - \nabla_i v_j - \nabla_j v_i
, \label{metretr} \\
\delta \phi & \rightarrow & \delta \phi - v^k\partial_k \delta \phi . \label{scetr}
\end{eqnarray}
Invariance of the wave function $\psi[\gamma_{ij}, \delta \phi]$ then leads to the condition
\be
\label{condrewf}
\int d^3 x v^j \partial_{x^i} \langle T_{ij}(\vect{x}) \hat{O}(\vect{y_1}) \hat{O}(\vect{y_2}) \cdots \hat{O}(\vect{y_n})\rangle=- 
 \sum_i \langle\hat{O}(\vect{y_1}) \cdots 
\delta \hat{O}(\vect{y_i}) \cdots  \hat{O}(\vect{y_n})\rangle,
\ee
where $\hat{O}$ is a schematic notation standing for both $T_{ij}, O,$ and $\delta \hat{O}(\vect{y_i})$ is the change 
in operator $\hat{O}(\vect{y_i})$ at the point $\vect{y_i}$. 
In particular when $\hat{O}=O$ is a scalar  we get for the three point
function
\be
\label{condnormtpt}
\begin{split}
\partial_{x^i} \langle T_{ij}(\vect{x}) O(\vect{y_1}) O(\vect{y_2})\rangle =
&[\partial_{x^j}\delta^3(\vect{x}-\vect{y_1})]\langle O(\vect{y_1}) O(\vect{y_2})\rangle \\
 &+ [\partial_{x^j} \delta^3(\vect{x}-\vect{y_2})]\langle O(\vect{y_1}) O(\vect{y_2})\rangle.
\end{split}
\ee

To summarize, the coefficient functions which arise in the wave function \eqref{wf1} satisfy all the symmetry properties of 
correlations functions involving a scalar operator of dimension $3$ and the stress energy tensor in a conformal field 
theory. Namely, they are invariant under the conformal symmetry group $SO(1,4)$ and satisfy the Ward identities due to 
conservation of the stress energy tensor. 

Let us end this section by noting that readers familiar with the AdS/CFT correspondence will hardly find the connection discussed
above between the coefficient functions and the correlation functions of a CFT surprising. 
For the   wave function in the inflationary context  \eqref{wf3} is the analogue of the bulk partition function in the AdS/CFT correspondence which in turn equals the CFT  partition function in the presence of sources. 

\section{Constraints of Conformal Invariance on the Correlation Function}
In this section we will discuss how the correlation function \eqref{defcorr}
 is constrained by the symmetries. 
This correlation function  is obtained from the  coefficient function, $\langle O O T_{ij}\rangle$, 
of the wave function in   \eqref{wf1}. 
We have argued in the previous section that as far as symmetries are concerned the coefficient functions  behave in exactly the same manner as corresponding correlation functions of a CFT. 
In our discussion below we will find it convenient to  adopt the language of CFT.
We remind the reader that this is only a kind of short-hand for   analyzing  the consequences of symmetries.
In particular,  we will not be  assuming any kind of  deeper dS/CFT type relation in our analysis. 

We will work in momentum space below. We derive our constraints in two
ways. The first is to directly act with the generators of conformal
transformations on the momentum space correlator. The other is to
translate the correlator into the spinor-helicity formalism, and then
use the conformal generators in terms of those variables. Of course,
we obtain the same differential equations with both approach. The
spinor-helicity formalism has the disadvantage of being a little more technical but leads to the
result a little more directly. The reader who is unfamiliar with  the
spinor-helicity formalism can skip \S3.2 and \S3.3 on a first reading and proceed from \S3.1 directly to \S4.

\noindent{\underline {\bf Notation:}}

Before proceeding let us list our conventions. We denote the three momentum by $\vect{k}$ below.
Its magnitude will be denoted simply by  $k \equiv |\vect{k}|$.  
Components will be denoted by $k_i, i=1, ...,3$ and indices will be
 raised 
and lowered by the flat space metric $\delta_{ij}$. 

\subsection{Direct Momentum Space Analysis}

In our conventions the momentum space scalar operator  is given by 
\be
\label{defmomop}
O(\vect{k}) \equiv \int d^3 x O(\vect{x}) e^{-i\vect{k}\cdot \vect{x}},
\ee
and similarly for $T_{ij}(\vect{k})$.
 
Translational 
 and  rotational invariance allows us to express the correlators in the form
\be
\label{formcorr} 
\begin{split}
\langle O(\vect{k}_1) O(\vect{k}_2) T_{ij}(\vect{k}_3)\rangle  &=  [k_{1i} k_{1j} f_1(k_1,k_2,k_3)+ k_{2i}k_{2j} f_1(k_2,k_1,k_3) \\
                             &+(k_{1i}k_{2j} + k_{2i} k_{1j}) f_2(k_1,k_2,k_3)  \\ 
                             &+ \delta_{ij} f_3(k_1,k_2,k_3)]
                             (2\pi)^3
                             \delta^3(\sum_i\vect{k}_i). 
\end{split}
\ee
 The overall delta function arises due to translational invariance. 
In the discussion below we will use $M_{ij}(\vect{k}_1,\vect{k}_2,\vect{k}_3)$ to denote the correlation function without
the overall delta function factor,
\be
\label{defM}
\langle O(\vect{k}_1) O(\vect{k}_2) T_{ij}(\vect{k}_3)\rangle =M_{ij}(\vect{k}_1,\vect{k}_2,\vect{k}_3) (2\pi)^3 \delta(\sum_i\vect{k}_i).
\ee

The three functions $f_1,f_2,f_3$ in \eqref{formcorr}
  at first sight could have also depended on inner products $\vect{k_1}\cdot \vect{k_2}$ etc. 
However using momentum conservation these can be expressed in terms of
the three scalars $k_i$. For example 
\be
\label{momcons1}
\vect{k_1}\cdot \vect{k_2} = {1\over 2} (k_3^2-k_1^2-k_2^2).
\ee
The correlator is symmetric under the exchange of $\vect{k}_1 \leftrightarrow \vect{k}_2$. As a result 
$f_2, f_3$ are symmetric under the exchange of their first two arguments. 
Since the operators  $O$ and $T_{ij}$ are dimension $3$ in position space and thus dimension $0$ in momentum space, 
scale invariance tells us that the $f_i$'s are dimension $1$. 

Next  we come to the non-trivial constraints due to special conformal transformations. 
The transformation in position space of the operators $O$ and $T_{ij}$ under an infinitesimal  special conformal 
transformation with  parameter $\eepsilon_i$ is given in \eqref{ospcft2} and \eqref{stressspcft2} respectively. 
In momentum space these take the form, 
\begin{eqnarray} 
\delta O(\vect{k}) & = & -\tilde{D} O(\vect{k}) \label{trspmo} , \\
\delta T_{ij}(\vect{k}) & = & 2 \tilde{M}^l_i T_{lj}+ 2 \tilde{M}^l_j T_{il} -\tilde{D} T_{ij} \label{trspt} , \\
\tilde{M}^l_i & \equiv & \eepsilon^l \partial_{k^i}-\eepsilon^i\partial_{k^l} \label{deftm} , \\
\tilde{D} & \equiv & \dotp[\eepsilon,k] \partial_{k^i} \partial_{k^i} - 2 k_j\partial_{k_j} (\vect{\eepsilon}\cdot \vect{\partial_k}).
\end{eqnarray}
These expressions agree with eq.(4.12) in \cite{Maldacena:2011nz}
 and in fact we have chosen essentially the same conventions to try and ensure 
readability. 

The condition for invariance of the correlator is 
\be
\label{inspctr}
\langle \delta O(\vect{k}_1) O(\vect{k}_2) T_{ij}(\vect{k}_3)\rangle + \langle O(\vect{k}_1) \delta O(\vect{k}_2) T_{ij}(\vect{k}_3)\rangle 
+ \langle O(\vect{k}_1) O(\vect{k}_2) \delta T_{ij}(\vect{k}_3)\rangle=0.
\ee
As  was argued in \cite{Maldacena:2011nz}  all terms involving derivatives that act on the  overall momentum conserving delta 
function sum to zero so  we will henceforth neglect the effect of the derivative operators acting on the delta function. 

Defining the operator 
\be
\label{deftheta}
\Theta (k)\equiv -{2\over k} {\partial \over \partial k} + {\partial^2 \over \partial k^2}  ,
\ee
where $k \equiv |\vect{k}|$ 
one can then show after some algebra that 
\be
\label{delo1}
\begin{split}
&\langle\delta O(\vect{k}_1) O(\vect{k}_2) T_{ij}(\vect{k}_3)\rangle  =  -2 
\dotp[\eepsilon, k_1] \delta_{ij} f_1 
+ 2(\eepsilon_i k_{1j} +\eepsilon_j k_{1i}) (1+ k_1 \partial_{k_1}) f_1 
\\ &+ 2(\eepsilon_i k_{2j}+\eepsilon_j k_{2i}) k_1 \partial_{k_1} f_2  +  
\dotp[\eepsilon,k_1] \Theta(k_1) [f_1 k_{1i} k_{1j} + f_1^T k_{2i} k_{2j} + f_2 (k_{1i} k_{2j} + k_{2i} k_{1j} ) + f_3 \delta_{ij}].
\end{split}
\ee
Here we have omitted  the overall delta function. 
We have also  introduced the notation
\be
\label{defft}
f_1^T(k_1,k_2,k_3) \equiv f_1(k_2,k_1,k_3).
\ee

At this stage it is useful to contract the LHS of \eqref{delo1}  with the symmetric (real) 
 polarization tensor $e^s_{ij}$ which is traceless and transverse to $\vect{k}_3$,
\be
\label{tratra}
e^{s,i}_i=e^s_{ij}k_3^i=0.
\ee
The $s$ here indicates that there are two possible choices for this tensor.
 This gives 
\begin{eqnarray}
\langle\delta O(\vect{k}_1) O(\vect{k}_2) T_{ij}(\vect{k}_3)\rangle e^{s,ij}& = & 4 b_i k_{1j} e^{s,ij} 
[(1+ k_1\partial_{k_1}) f_1 - k_1  \partial_{k_1} f_2] \nonumber \\
&&  + \dotp[b,k_1] \Theta(k_1) (2 f_2 - f_1 -f_1^T) k_{1i} k_{2j} e^{s,ij},  \label{aftcon}
\end{eqnarray} 
where we have used the condition
\be
\label{condepsa}
e^{s,ij}k_{1i}=-e^{s,ij}k_{2i}=0.
\ee

Similarly we get 
\be
 \label{aftcontwo}
\begin{split}
\langle O(\vect{k}_1) \delta O(\vect{k}_2) T_{ij}(\vect{k}_3)\rangle e^{s,ij} = & -4 b_i k_{1j} e^{s,ij} 
[(1+ k_2\partial_{k_2}) f_1^T - k_2  \partial_{k_2} f_2]  \\
&+ \dotp[b,k_2] \Theta(k_2) (2 f_2 - f_1 -f_1^T) k_{1i} k_{2j}
e^{s,ij}.
\end{split}
\ee

And also 
\be
\label{aftconthree}
\begin{split}
\langle O(\vect{k}_1) O(\vect{k}_2) \delta T_{ij}(\vect{k}_3)\rangle e^{s,ij}  = & -{4 \over k_3}  b_i k_{1j} e^{s,ij} [(\vect{k_3}\cdot \vect{k_1}) \partial_{k_3} (f_1-f_2) -(\vect{k_3}\cdot \vect{k_2}) \partial_{k_3}(f_1^T-f_2)] \\
&+ \vect{b} \cdot \vect{k_3} \Theta (k_3) (2 f_2 - f_1 -f_1^T).  
\end{split}
\ee
Adding \eqref{aftcon}, \eqref{aftcontwo} and \eqref{aftconthree} and setting the total change  to vanish 
finally gives the equation
\be
\label{finalspc}
\begin{split}
4 b_i k_{1j} e^{s,ij} \Big[&(1+ k_1\partial_{k_1}) f_1
-(1+k_2 \partial_{k_2}) f_1^T
+(k_2 \partial_{k_2}-k_1 \partial_{k_1})f_2 \\
&- {(\vect{k_3}\cdot \vect{k_1}) \over
   k_3} \partial_{k_3}(f_1-f_2)  + {\dotp[k_3, k_2] \over
   k_3} \partial_{k_3}(f_1^T-f_2) \Big]  \\
+ k_{1i} k_{2j} e^{s,ij} \Big[&\dotp[b,k_{1}] \Theta(k_1) + \dotp[b,k_2] \Theta(k_2) + \dotp[b,k_3] \Theta(k_3)\Big]
(2f_2-f_1-f_1^T)  =0. 
\end{split}
\ee
This is the main equation we will use to derive the constraints imposed by the special conformal transformations. 
 
There are three linearly independent values that $\vect{b}$ can take in \eqref{finalspc}. 
Choosing $\vect{b} \propto \vect{k}_3$ gives

\be
\label{eq1spc}
[\vect{k}_3\cdot \vect{k}_1 \Theta(k_1) + \dotp[k_3, k_2]\Theta(k_2) + k_3^2 \Theta(k_3)] 
S(k_1,k_2,k_3)=0,
\ee
where 
\be
\label{defS}
S(k_1,k_2,k_3)={1\over 2}[f_1(k_1,k_2,k_3) + f_1(k_2,k_1,k_3)-2 f_2(k_1,k_2,k_3)].
\ee 

Choosing  $\vect{b}\propto \vect{k}_{1\perp } = \vect{k}_1-\vect{k}_3 { \dotp[k_1, k_3] \over k_3^2}$
gives
\begin{eqnarray}
4[{-\vect{k_2}\cdot \vect{k_3} \over k_3^2} k_1\partial_{k_1} S + {\vect{k_1}\cdot \vect{k_3} \over k_3^2} k_2 \partial_{k_2} S 
-{(k_1^2-k_2^2)\over k_3^2} S + {3\over 2} {(k_1^3-k_2^3) \over k_3^2}] && \nonumber \\ 
- (k_1^2 - {(\vect{k_3}\cdot \vect{k_1})^2 \over k_3^2}) (\Theta(k_1) -\Theta(k_2))S &=&0, \label{eq2spc}
\end{eqnarray} 
as shown in Appendix \ref{appendix2}. The term inhomogeneous in $S$
above arises due to the use of  the Ward identity for conservation of
the stress tensor.
We take the two-point function of the scalar $O(k)$ to be  normalized so that
\be
\label{tptO}
\langle O(k_1) O(k_2) \rangle=(2\pi)^3 \delta(\vect{k}_1+\vect{k}_2) |\vect{k}_1|^3.
\ee
 The  Ward identity for  conservation of the stress tensor, \eqref{condnormtpt}  then takes the form 
\be
\label{Ward}
M_{ij} k_3^j=-k_1^3 k_1^j-k_2^3 k_2^j,
\ee
where $M_{ij}$ is defined in \eqref{defM}. 

Finally we can choose $\vect{b}$ to be orthogonal to all the $\vect{k}_i$'s so that $\vect{b}\cdot \vect{k}_i=0$. 
For a suitable choice of polarization $b_i k_{1j} e^{s,ij}$ will not vanish and as discussed in Appendix \ref{appendix2}  
\eqref{finalspc}
then becomes  
\begin{eqnarray}
-\dotp[k_2,k_3] k_1\partial_{k_1} S + \dotp[k_1, k_3] k_2 \partial_{k_2} S 
-(k_1^2-k_2^2) S  
+ {3\over 2} (k_1^3-k_2^3) && =0 \label{eq3spc}.
\end{eqnarray}

Subtracting  \eqref{eq2spc} and \eqref{eq3spc} then gives 
\be
\label{final1}
(\Theta(k_1) -\Theta(k_2))S=0.
\ee
Substituting this in \eqref{eq1spc} then gives
\be
\label{final2}
(\Theta(k_1) -\Theta(k_3))S = 0.
\ee
Equations \eqref{eq3spc}, \eqref{final1} and \eqref{final2} can be taken to be the three final equations which arise because
of special conformal invariance. 

Before proceeding  let us note here that 
from \eqref{formcorr}, \eqref{condepsa}  and \eqref{defS}  we get that
\be
\label{fsola}
\langle O(k_1) O(k_2) T_{ij}(k_3)\rangle e^{s,ij}=-2 (2\pi)^3 \delta(\sum_i\vect{k}_i)e^{s,ij} k_{1i}k_{2j}  S,
\ee
where $e^{s}_{i j}$ is a  traceless polarization tensor transverse to
$\vect{k}_3$.

\subsection{Analysis using the Spinor Helicity Formalism \label{secspinhelconf}}
We now rederive these differential equations in a second way, using the spinor helicity
formalism of \cite{Maldacena:2011nz} and \cite{Raju:2012zs}. Our
notation is described in detail in Appendix \ref{appspinor}.

Although the correlator in \eqref{defM} appears to have several
independent components, the use of the Ward identities for the
conservation of the stress-tensor and its tracelessness, leave us with
only two components. We can extract both of these by considering the quantities:
\be
\begin{split}
&M^{+}(\vect{k_1}, \vect{k_2}, \vect{k_3}) (2 \pi)^3 \delta(\vect{k_1}
+ \vect{k_2} + \vect{k_3}) ={1 \over \norm{k_1} \norm{k_2} \norm{k_3}} e^{+}_{i j} \langle O(\vect{k_1}) O(\vect{k_2}) T^{i
  j}(\vect{k_3}) \rangle, \\
& M^{-}((\vect{k_1}, \vect{k_2},
\vect{k_3}) (2 \pi)^3 \delta(\vect{k_1}
+ \vect{k_2} + \vect{k_3}) = {1 \over \norm{k_1} \norm{k_2}
  \norm{k_3}}  e^{-}_{i j} \langle O(\vect{k_1})
O(\vect{k_2}) T^{i j}(\vect{k_3}) \rangle,
\end{split}
\ee
where  $e^{+}$ and $e^{-}$ are symmetric traceless tensors that are
transverse to $\vect{k_3}$. We caution the reader that these are
linear combinations of the real polarization tensors $e^s$ that have
appeared previously and whenever we use these ``circularly
polarized'' tensors, we put a $\pm$ rather than a $s$ in the
superscript. We give explicit expressions for these
tensors in Appendix \ref{appspinor}. The pre-factor of ${1 \over \norm{k_1} \norm{k_2}
  \norm{k_3}}$ is included for convenience. 

The momentum space correlators manifestly have an $SO(3)$ symmetry and just this
allows us to write
\be
\label{spinorhelans}
M^{-} (\vect{k_1}, \vect{k_2}, \vect{k_3}) =  \tilde{R}(\norm{k_1},
\norm{k_2}, \norm{k_3}) {\dotl[\la_3, \la_1]^2 \dotl[\la_3, \la_2]^2
  \over \dotl[\la_1, \la_2]^2}.
\ee
We can write a similar expression for $M^{+}$, but this leads to the
same constraints, and our analysis can be performed entirely with the
expression above. We now need to derive constraints on the function $\tilde{R}$, which
depends just on the norms of the momenta.

The constraints of special conformal invariance, in the spinor
helicity formalism (see the Appendix for a derivation) can be written as
\be
\label{finalconstraint}
b_i \sigma^i_{\alpha \dot{\alpha}} \sum {\partial \over \partial \la_{n \alpha}} {\partial \over \partial \lb_{n \dot{\alpha}}} M^{-}  = \left({\dotp[b,k_1] \over \norm{k_1}^2}+ {\dotp[b,k_2] \over \norm{k_2}^2} \right) M^{-} + W,
\ee
where $W$ is the Ward identity term
\be
\label{Wdef}
W = {3 b^k e^{-}_{k j} k_{3 i}  \over \norm{k_3}^3 } \langle O(\vect{k_1}) O(\vect{k_2}) T^{i j} (\vect{k_3})\rangle + (i \leftrightarrow j).
\ee

Now, notice that
\be
\begin{split}
b_i \sigma^i_{\alpha \dot{\alpha}} {\partial \over \partial \la_{1
    \alpha}} {\partial \over \partial \lb_{1 \dot{\alpha}}} M^{-} = &b_i
\sigma^i_{\alpha \dot{\alpha}} {\dotl[\la_3, \la_1]^2 \dotl[\la_3,
  \la_2]^2 \over \dotl[\la_1, \la_2]^2} {\partial \over \partial
  \la_{1 \alpha}} {\partial \over \partial \lb_{1 \dot{\alpha}}}
\tilde{R}  \\ &+ b_i \sigma^i_{\alpha \dot{\alpha}} \left({\partial
    \over \partial \lb_{1 \dot{\alpha}}} \tilde{R}\right)  {\partial
  \over \partial \la_{1 \alpha}} {\dotl[\la_3, \la_1]^2 \dotl[\la_3,
  \la_2]^2 \over \dotl[\la_1, \la_2]^2}.
\end{split}
\ee
We see that 
\be
\begin{split}
 {\partial \over \partial \la_{1 \alpha}} {\dotl[\la_3, \la_1]^2 \dotl[\la_3, \la_2]^2 \over \dotl[\la_1, \la_2]^2} &= \dotl[\la_3, \la_2]^2 \left({2 \dotl[\la_1, \la_3] \la_3^{\alpha} \over \dotl[\la_1, \la_2]^2} - 2{ \dotl[\la_1, \la_3]^2\la_2^{\alpha} \over \dotl[\la_1, \la_2]^3}\right) \\
&= {2 \dotl[\la_3, \la_2]^3 \dotl[\la_1, \la_3] \over \dotl[\la_1, \la_2]^3} \la_1^{\alpha},
\end{split}
\ee
where we have used the Schouten identity in the last step. Also,
\be
{\partial \tilde{R} \over \partial \lb_{1 \dot{\alpha}}} = {\partial \tilde{R} \over \partial \norm{k_1}} {\partial \norm{k_1} \over \partial \lb_{1 \dot{\alpha}}} = {1 \over 2} \bar{\sigma}_0^{\dot{\alpha} \beta} \la_{1 \beta} {\partial \tilde{R} \over \partial \norm{k_1}}.
\ee
This tells us that
\be
b_i \sigma^i_{\alpha \dot{\alpha}} \left({\partial \over \partial \lb_{1 \dot{\alpha}}} \tilde{R}\right)  {\partial \over \partial \la_{1 \alpha}} {\dotl[\la_3, \la_1]^2 \dotl[\la_3, \la_2]^2 \over \dotl[\la_1, \la_2]^2} =   2  { \dotl[\la_3, \la_2]^3 \dotl[\la_1, \la_3] \over \dotl[\la_1, \la_2]^3}  \norm{k_1} \dotp[b, \epsilon_1] {\partial \tilde{R} \over \partial \norm{k_1}},
\ee
where $\epsilon_1$ is the transverse and null ``polarization vector''
defined by \eqref{polarizationvects2}.

Putting this algebra together, we see that 
\be
\label{firstparticleaction}
b_i \sigma^i_{\alpha \dot{\alpha}} {\partial \over \partial \la_{1 \alpha}} {\partial \over \partial \lb_{1 \dot{\alpha}}} M^{-} = {\dotp[b, k_1] \over 2} {\dotl[\la_3, \la_1]^2 \dotl[\la_3, \la_2]^2 \over \dotl[\la_1, \la_2]^2}  { {\partial^2 \over \partial \norm{k_1}^2}}   \tilde{R}  + 2  { \dotl[\la_3, \la_2]^3 \dotl[\la_1, \la_3] \over \dotl[\la_1, \la_2]^3}  \norm{k_1} \dotp[b,\epsilon_1] {\partial \tilde{R} \over \partial \norm{k_1}}.
\ee
By interchanging $1 \leftrightarrow 2$, we see that 
\be
\label{secondparticleaction}
b_i \sigma^i_{\alpha \dot{\alpha}} {\partial \over \partial \la_{2 \alpha}} {\partial \over \partial \lb_{2 \dot{\alpha}}} M^{-} = {\dotp[b,k_2] \over 2} {\dotl[\la_3, \la_2]^2 \dotl[\la_3, \la_1]^2 \over \dotl[\la_2, \la_1]^2}  { {\partial^2 \over \partial \norm{k_2}^2}}   \tilde{R}  + 2  { \dotl[\la_3, \la_1]^3 \dotl[\la_2, \la_3] \over \dotl[\la_2, \la_1]^3}  \norm{k_2} \dotp[b,\epsilon_2] {\partial \tilde{R} \over \partial \norm{k_2}}.
\ee

Turning to the third particle, we note that
\be
\begin{split}
 {\partial \over \partial \la_{3 \alpha}} {\dotl[\la_3, \la_1]^2 \dotl[\la_3, \la_2]^2 \over \dotl[\la_1, \la_2]^2} &= 2 {\dotl[\la_3, \la_1] \dotl[\la_3, \la_2] \over \dotl[\la_1, \la_2]} \left( \la_1^{\alpha} {\dotl[\la_3, \la_2] \over \dotl[\la_1, \la_2]} + \la_2^{\alpha} {\dotl[\la_3, \la_1] \over \dotl[\la_1, \la_2]} \right) \\ &\equiv  2 {\dotl[\la_3, \la_1] \dotl[\la_3, \la_2] \over \dotl[\la_1, \la_2]} \la_4^{\alpha},
\end{split}
\ee
where we have defined a new spinor $\la_4$ in the last step for convenience. This leads to
\be
\label{thirdparticleaction}
b_i \sigma^i_{\alpha \dot{\alpha}} {\partial \over \partial \la_{3 \alpha}} {\partial \over \partial \lb_{3 \dot{\alpha}}} M^{-} = {\dotp[b,k_3] \over 2} {\dotl[\la_3, \la_2]^2 \dotl[\la_3, \la_1]^2 \over \dotl[\la_2, \la_1]^2}  { {\partial^2 \over \partial \norm{k_3}^2}}   \tilde{R} +  2 {\dotl[\la_3, \la_1] \dotl[\la_3, \la_2] \over \dotl[\la_1, \la_2]} (b_{\alpha \dot{\alpha}} \lad_3^{\dot{\alpha}} \la_4^{\alpha}) {\partial \tilde{R} \over \partial \norm{k_3}},
\ee
where recall that $b_{\alpha \dot{\alpha}} = b_i \sigma^i_{\alpha \dot{\alpha}}$, where the sum on $i$ runs only over $0,1,2$.

\subsubsection{Equations for Conformal Invariance}
 The choice of $\vect{b}$ allows us to project this in various directions. 
It is most convenient to take $\vect{b} = \vect{\ep_n}$, with $n = 1, 2, 3$ in turn. 
\paragraph{$\boldsymbol{b \propto \ep_3}$:}
Let us start with $b_{\alpha \dot{\alpha}} = 2 \la_{3 \alpha} \lad_{3 \dot{\alpha}} = 2  \norm{k_3} \ep_3$. The advantage of this particular case is that the Ward identity term does not contribute
for this choice of $\vect{b}$ and moreover, \eqref{thirdparticleaction} drops out since $\vect{\ep_3} \cdot \vect{k_3} = \vect{\ep_3} \cdot \vect{\ep_4} = 0$.  With this choice of $\vect{b}$, we see that
\be
\label{bchoice1}
\begin{split}
&\norm{k_1} \vect{b} \cdot \vect{\ep_1} =  \dotl[\la_3, \la_1]^2, \quad \norm{k_2} \vect{b} \cdot \vect{\ep_2} =  \dotl[\la_3, \la_2]^2, \\
&\vect{b} \cdot \vect{k_1} = \dotl[\la_3, \la_1] \dotlm[\la_3, \lb_1] =  {\dotl[\la_3, \la_1] \dotl[\la_3, \la_2] \over \dotl[\la_2, \la_1]} \left(\norm{k_1} + \norm{k_2} - \norm{k_3} \right), \\ & \vect{b} \cdot \vect{k_2} =   {\dotl[\la_3, \la_1] \dotl[\la_3, \la_2] \over \dotl[\la_1, \la_2]} \left(\norm{k_1} + \norm{k_2} - \norm{k_3} \right).
\end{split}
\ee

Adding \eqref{firstparticleaction} and \eqref{secondparticleaction} and substituting \eqref{bchoice1}, we see that $\tilde{R}$ must 
satisfy the equation:
\be
\label{conformaldiffeq}
 {1 \over 2} \left({\partial^2 \tilde{R} \over \partial \norm{k_1}^2} - {\partial^2 \tilde{R} \over \partial \norm{k_2}^2}\right) - {2 \over \norm{k_1} + \norm{k_2} - \norm{k_3} } \left( {\partial \tilde{R} \over \partial \norm{k_1}} - {\partial \tilde{R} \over \partial \norm{k_2}} \right) = \tilde{R} \left({1 \over \norm{k_1}^2} - {1 \over \norm{k_2}^2} \right)
\ee

\paragraph{$\boldsymbol{b \propto \ep_1}$:}
Now, let us choose $b_{\alpha \dot{\alpha}} = 2 \la_{1 \alpha} \lad_{1 \dot{\alpha}} = 2 \norm{k_1} \ep_1$.
We need the Ward identity term. In position space, this is given by
\eqref{condnormtpt} and its Fourier transform is given by
\eqref{Ward}. 
Using this, and keeping track of the factor of ${1 \over 2}$ in
\eqref{dotproductspinor}, we see that \eqref{Wdef} evaluates to:
\be
\begin{split}
W &= 3 {\dotl[\la_1, \la_3]^2  \over \norm{k_3}^5 \norm{k_1} \norm{k_2}} \dotl[\la_3, \la_1] \dotlm[\la_3, \lb_1] \left(\norm{k_1}^3 - \norm{k_2}^3 \right) \\ &=  3  {\dotl[\la_3, \la_1]^3 \dotl[\la_3, \la_2] \over \norm{k_3}^5 \norm{k_1} \norm{k_2} \dotl[\la_2, \la_1]} \left(\norm{k_1} + \norm{k_2} - \norm{k_3} \right) \left(\norm{k_1}^3 - \norm{k_2}^3 \right).
\end{split}
\ee

Next we see that
\be
\begin{split}
&\vect{b} \cdot \vect{k_2} =  {\dotl[\la_1, \la_2] \dotl[\la_1, \la_3] \over \dotl[\la_3, \la_2]} \left( \norm{k_3} + \norm{k_2}  - \norm{k_1} \right) = -\vect{b} \cdot \vect{k_3} \\
&b_{\alpha \dot{\alpha}} \la_4^{\alpha} \lad_3^{\dot{\alpha}} =  -\dotl[\la_1, \la_3]^2, \quad \norm{k_2} \vect{b} \cdot \vect{\ep_2} = \dotl[\la_1, \la_2]^2.
\end{split}
\ee

This leads to the following equation for $\tilde{R}$:
\be
\label{eqe1}
\begin{split}
&{1 \over 2} \left({\partial^2 \tilde{R} \over \partial \norm{k_2}^2} - {\partial^2 \tilde{R} \over \partial \norm{k_3}^2}\right) + {2 \over \norm{k_2} + \norm{k_3} - \norm{k_1} } \left( {\partial \tilde{R} \over \partial \norm{k_2}} - {\partial \tilde{R} \over \partial \norm{k_3}} \right) \\
&= {\tilde{R} \over \norm{k_2}^2} + 3 {\norm{k_1} + \norm{k_2} - \norm{k_3} \over \norm{k_2} + \norm{k_3} - \norm{k_1}} {\left(\norm{k_1}^3 - \norm{k_2}^3 \right) \over \norm{k_3}^5 \norm{k_1} \norm{k_2}}.
\end{split}
\ee

\paragraph{$\boldsymbol{b \propto \ep_2}$:}
 We do not need to explicitly 
compute the term with  $b \propto \ep_2$ after this, since that equation should be obtainable just by interchanging particles $1$ and $2$. So, we can immediately see that we must have the equation:
\be
\label{eqe2}
\begin{split}
& {1 \over 2} \left({\partial^2 \tilde{R} \over \partial \norm{k_1}^2} - {\partial^2 \tilde{R} \over \partial \norm{k_3}^2}\right) + {2 \over \norm{k_1} + \norm{k_3} - \norm{k_2} } \left( {\partial \tilde{R} \over \partial \norm{k_1}} - {\partial \tilde{R} \over \partial \norm{k_3}} \right) \\ &= {\tilde{R} \over \norm{k_1}^2} + 3 {\norm{k_2} + \norm{k_1} - \norm{k_3} \over \norm{k_1} + \norm{k_3} - \norm{k_2}} {\left(\norm{k_2}^3 - \norm{k_1}^3 \right) \over \norm{k_3}^5 \norm{k_2} \norm{k_1}}.
\end{split}
\ee

It is more convenient to derive another homogeneous equation by combining \eqref{eqe2} and \eqref{eqe1}. From these two, we get:
\be
\label{eqcombinedhomog}
\begin{split}
&(\norm{k_{2}} + \norm{k_{3}} - \norm{k_{1}}) {\partial^2 \tilde{R} \over \partial \norm{k_{2}}^2} + (\norm{k_{1}} + \norm{k_{3}} - \norm{k_{2}}) {\partial^2 \tilde{R} \over \partial \norm{k_{1}}^2} - 2 \norm{k_{3}} {\partial^2 \tilde{R} \over \partial \norm{k_{3}}^2} \\ &+ 2 \left({\partial \tilde{R} \over \partial \norm{k_{2}}} + {\partial \tilde{R} \over \partial \norm{k_{1}}} - 2 {\partial \tilde{R} \over \partial \norm{k_{3}}} \right) = \tilde{R} \left({\norm{k_{1}} + \norm{k_{3}} - \norm{k_{2}} \over \norm{k_{1}}^2} + {\norm{k_{2}} + \norm{k_{3}} - \norm{k_{1}} \over \norm{k_{2}}^2} \right).
\end{split}
\ee

\subsection{Comparison with direct momentum space computations}
It is now useful to translate our notation back to that of the
previous subsection and write down our final equations. Contracting
\eqref{formcorr} with the polarization tensor and using spinor
identities to rewrite the answer in the form \eqref{spinorhelans}
leads to a relation between $\tilde{R}$ and $S$ 
\be
\label{reltilders}
\tilde{R} =  -{(\norm{k_1} + \norm{k_2} - \norm{k_3})^2 \over
  \norm{k_1} \norm{k_2} \norm{k_3}^3} {S \over 2}.
\ee
(See Appendix \ref{tildeRS} for a derivation).
From here, \eqref{conformaldiffeq} tells us that $S$ must satisfy the equation:
\be
\label{theta12eq}
\left(\Theta(\norm{k_1}) - \Theta(\norm{k_2})\right) S = 0.
\ee
This is precisely what we would get by substituting $\vect{b} = \vect{\ep_3}$ in \eqref{finalspc}.

We can also write \eqref{eqcombinedhomog} in terms of $S$ rather than $\tilde{R}$. When we do this, we find that $S$ must satisfy the equation:
\be
\begin{split}
&8 {k_{1}} {k_{2}} \left[S - k_1 {\partial S \over \partial k_1} - k_2
  {\partial S \over \partial k_2} - k_3 {\partial S \over \partial
    k_3} \right] + 2 k_1 k_2 k_3 (k_1 + k_2 - k_3) \Theta(k_3) S \\ & + k_1 k_2 \left( (k_1 - k_3)^2 - k_2^2 \right) \Theta(k_2) S + k_1 k_2 \left(-k_1^2 + (k_2 - k_3)^2 \right) \Theta(k_1) S  = 0.
\end{split}
\ee
If we now use the fact that $S$ has dimension $1$ and also the equation \eqref{theta12eq}, we find the remarkably simple equation:
\be
\label{theta13eq}
\left(\Theta(\norm{k_1}) - \Theta(\norm{k_3})\right) S = 0.
\ee

So \eqref{theta13eq} and \eqref{theta12eq} are our final homogeneous
equations, which can also be obtained directly in momentum
space. These are separate from the inhomogeneous equation
\eqref{eqe1}. The inhomogeneous equation can also be shown to be
equivalent to \eqref{eq2spc}. By substituting \eqref{reltilders} into \eqref{eqe1}, we
find that
\be
\begin{split}
&2 {k_1} {k_3} \left(3 {k_1}^2-3 {k_2}^2+2 {k_2}
   {k_3}+{k_3}^2\right)
   {\partial S \over \partial \norm{k_3}}+2 {k_3}^2 \left(3
   {k_1}^2+({k_2}-{k_3})^2\right)
   {\partial S \over \partial \norm{k_1}}\\ & + k_3^2 {k_1}
   ({k_1}+{k_2}-{k_3}) ({k_1}-{k_2}+{k_3}) \left(-
  {\partial^2 S \over \partial \norm{k_3}^2}+
   {\partial^2 S \over \partial \norm{k_1}^2} \right)-4 {k_1} \left(3
   {k_1}^2-3 {k_2}^2+{k_3}^2\right)
   S \\ &+12 \left({k_1}^4-{k_1}
   {k_2}^3\right) = 0.
\end{split}
\ee
Now, if we write ${\partial^2 S \over \partial k_3^2} =
\Theta(k_3) S + {2 \over k_3} {\partial S \over \partial k_3}$, use
the fact that $\left[\Theta(k_3) - \Theta(k_1)\right] S = 0$, and
collect the terms proportional to the different partial derivatives of
$S$, we find that this reduces to:
\be
\begin{split}
& {k_3} \left( {k_1}^2- {k_2}^2+ {k_3}^2 \right)
   {\partial S \over \partial \norm{k_3}}+2 {k_3}^2 k_1
   {\partial S \over \partial \norm{k_1}}- \left(3
   {k_1}^2-3 {k_2}^2+{k_3}^2\right)
   S +3 \left({k_1}^3- {k_2}^3\right) = 0.
\end{split}
\ee
Now, if we use the fact that 
\be
\sum_m k_m {\partial S \over \partial k_m}= S,
\ee
to substitute for $k_3 {\partial S \over \partial k_3}$ in the
equation above, and also use \eqref{momcons1} we find that
\be
\begin{split}
& -(\vect{k_1} \cdot \vect{k_3}) {k_2}  {\partial \over \partial k_2}
S + (\vect{k_2} \cdot \vect{k_3}) {k_1}  {\partial \over \partial k_1} S
   - \left(
   {k_1}^2- {k_2}^2\right)
   S +{3 \over 2} \left({k_1}^3- {k_2}^3\right) = 0.
\end{split}
\ee
This is exactly the same as \eqref{eq2spc}, if we use \eqref{theta12eq}.

This concludes our demonstration that the differential equations obtained in
spinor helicity variables are the same as those obtained directly in
momentum space.

\section{Solving the Conformal Constraints}
The  three point correlator involving two scalar and one tensor perturbations was calculated 
for a model of inflation in \cite{Maldacena:2002vr}. The answer is
given in equations (4.10) and (4.11) of 
\cite{Maldacena:2002vr} in terms of the function
\be
\label{defI}
I=-(k_1+k_2+k_3) + 
{\sum_{i>j} k_i k_j \over (k_1+k_2+k_3)}+ {k_1k_2k_3\over (k_1+k_2+k_3)^2}.
\ee
From this result  we can read off the functional form for the 
corresponding $ \langle O O T_{ij} \rangle$ coefficient.  This gives 
\be
\label{valsm}
S=- I = - [-(k_1+k_2+k_3) + {\sum_{i>j}k_i k_j \over (k_1+k_2+k_3)}+{k_1k_2k_3\over (k_1+k_2+k_3)^2}].
\ee
It is easy to check that this function solves the three equations \eqref{eq2spc}, \eqref{final1}, \eqref{final2} above. 

\subsection{Uniqueness}
In this subsection we will see that \eqref{valsm} is the  unique solution to \eqref{eq3spc}, \eqref{final1}, \eqref{final2}
 which meets all the required conditions.

We begin by noting that the set of functions 
\be
f_{z}(k) = (1 + i k z)e^{-i k z},
\ee
with $z$ allowed to range over both positive and negative values forms a complete set. 
Any function ${\cal H}(k)$ can be expanded in terms of this set,
\be
\label{expcalh1}
{\cal H}(k)=\int_{-\infty}^{\infty}\tilde{\phi}(z) f_z(k) dz.
\ee

The point is that $\tilde{\phi}$ is a kind of souped up Fourier transform of ${\cal H}(k)$.
Let $\phi(k)$ be the Fourier transform of $\tilde{\phi}(z)$. Then
\eqref{expcalh1} gives 
\be
\label{expcalh}
{\cal H}(k) =  {\phi}(k) - k {\phi}'(k) = -k^2 {d \over d k} \left( {\phi(k) \over k} \right),
\ee
which  can be solved to obtain  
\be
\label{detphi}
\phi(k) = -k \int^k {{\cal H}(x) \over x^2} d x, 
\ee
and correspondingly
\be
\tilde{\phi}(z) = \int_{-\infty}^{\infty} - \left[k e^{i k z}  \int^k {{\cal H}(x) \over x^2} d x  \right] {d k \over 2 \pi}. 
\ee
Note that \eqref{detphi} determines $\phi(k)$ up to a term proportional to $k$ and this in turns leads to an ambiguity 
  proportional
to $\delta'(z)$ in $\tilde{\phi}(z)$, but this ambiguity drops out of the integral in \eqref{expcalh1}  leading to
a well defined value for ${\cal H}(k)$. 
 
Thus the most general solution can be expanded as 
\be
\label{completesol}
\begin{split}
S(k_1, k_2, k_3) = \int \Big[&(1 + i \norm{k_1} z_1) e^{-i \norm{k_1} z_1} (1 + i \norm{k_2} z_2) e^{-i \norm{k_2} z_2} \\ &\times (1 + i \norm{k_3} z_3) e^{-i \norm{k_3} z_3} {\cal M}(z_1, z_2, z_3) \Big] d z_1 d z_2 dz_3 ,
\end{split}
\ee
where each $z_i$ integral runs over $(-\infty, \infty)$.

Now note that since 
\be
\label{evaltheta}
\Theta(k) f_z(k) = -z^2 f_z(k),
\ee
the functions $f_z(k)$ are eigenvectors of the operator $\Theta(k)$.\footnote{The functions 
$f_z(k)$ are in fact  solutions to the massless scalar equation in de
Sitter space with $z$ being conformal time.}
It then follows that \eqref{final1}, \eqref{final2}, for $S$ given in \eqref{completesol}
  lead to the conditions 
\be
\label{condz}
z_1^2=z_2^2=z_3^3.
\ee
As a result an allowed solution 
can be written in the following form:
\be
S = \sum_{n_1, n_2, n_3 = \pm 1} \int_0^{\infty} {\cal F}_{n_1 n_2 n3}(z) {\cal M}_{n_1 n_2 n_3}(z) d z,
\ee
where ${\cal M}_{n_1, n_2, n_3}$ are a  set of 8 functions for the 8 possible combinations of $n_1, n_2, n_3$ and
\be
{\cal F}_{n_1 n_2 n_3}(z) = (1 + i n_1 \norm{k_1} z) e^{-i n_1 \norm{k_1} z} 
(1 + i n_2 \norm{k_2} z) e^{-i n_2 \norm{k_2} z} (1 + i n_3 \norm{k_3} z) e^{-i n_3 \norm{k_3} z}.
\ee
Next, we  apply the dilatation constraint:
\be
\left(k_1 { \partial \over \partial k_1}  + k_2 { \partial \over \partial k_2}  + k_3 {\partial \over \partial k_3} \right) S =S .
\ee
We notice that:
\begin{align}
\label{mintorig} 
&\left( {k_1 \partial \over \partial k_1}  + {k_2 \partial \over \partial k_2}  + {k_3 \partial \over \partial k_3} \right) S - S 
= \sum_{n_1,n_2,n_3=\pm 1} \int_{0}^{\infty}  {\cal M}_{n_1 n_2 n_3}(z)  \left(z {\partial \over \partial z} - 1 \right) {\cal F}_{n_1 n_2 n_3}(z)  \\
\label{mintbyparts} &= -\sum_{n_1,n_2,n_3=\pm 1}\int_0^{\infty} \left( {\partial \over \partial z} z  + 1 \right) {\cal M}_{n_1 n_2 n_3}(z) {\cal F}_{n_1 n_2 n_3}(z) dz,
\end{align}
which leads to
\be
-{\partial \over \partial z} z  {\cal M}_{n_1, n_2 n_3}(z) = {\cal M}_{n_1, n_2, n_3}(z).
\ee
This provides us with
\be
\label{measureans}
{\cal M}_{n_1, n_2, n_3} = {m_{n_1, n_2, n_3} \over z^2},
\ee
where $m_{n_1, n_2, n_3}$ is an arbitrary constant. Essentially all
that we are saying that the $z$ dependence of ${\cal M}_{n_1,n_2,n_3}$
is fixed by noting that it must have dimension $2$ and $z$ has
dimension $-1$.

In going from \eqref{mintorig} to \eqref{mintbyparts}, we tacitly assumed that ${\cal M}$ was regular at 
the origin so that we could  drop the boundary term at $0$. However, the result in 
 \eqref{measureans} makes \eqref{mintorig} divergent both at $0$ and at $\infty$.
We can be more careful as follows.
To define the integral at $z = \infty$, we can analytically continue the correlator to give the $\norm{k}_i$ a small imaginary part. To define the integral at $z = 0$, we can define it by:
\be
\label{Ilimitz0}
\begin{split}
&S =  \sum_{n_1, n_2, n_3 = \pm 1} m_{n_1 n_2 n_3} \int_0^{\infty}  {\cal F}_{n_1 n_2 n_3}(z) {d z \over z^2}
\equiv  \sum_{n_1, n_2, n_3 = \pm 1} m_{n_1 n_2 n_3} \left. \int_{\epsilon}^{\infty}  {\cal M}_{n_1 n_2 n_3}(z)
\right|_{\ep^0},
\end{split}
\ee
which means that we regulate the integral, by changing the range to $(\epsilon,\infty)$ and then pick up the 
$\epsilon^0$ term. This prescription now makes the resulting integral well defined
while  preserving its  behaviour under scale transformations.

The prescription above leads to:
\be
\begin{split}
S = \sum_{n_1, n_2, n_3 = \pm 1} m_{n_1 n_2 n_3} \Big(
 &-n_1 n_2 n_3 \frac{k_{2} k_{3}
   k_{1}}{(n_1 k_{1}+n_2 k_{2}+n_3k_{3})^2}+n_1  k_{1}+n_2 k_{2}+
   n_3 k_{3}\\
&-\frac{n_1 n_2 k_{1} k_{2}+ n_2 n_3 k_{3} k_{2}+ n_1 n_3 k_{1}
   k_{3}}{n_1 k_{1}+n_2 k_{2}+n_3 k_{3}} \Big).
\end{split}
\ee
Actually there are only four distinct terms in the sum above since the  function of $k_i$'s within the bracket on the RHS above
only changes by an overall sign when the sign of all three $n_i$'s is changed. 
We can use this property to fix $n_3=+1$ so that $S$ is given by a sum over four terms 
\be
\label{ftform}
\begin{split}
S = \sum_{n_1, n_2 = \pm 1} m_{n_1 n_2 } \Big(
 &-n_1 n_2  \frac{k_{2} k_{3}
   k_{1}}{(n_1 k_{1}+n_2 k_{2}+k_{3})^2}+n_1  k_{1}+n_2 k_{2}+
    k_{3}\\
&-\frac{n_1 n_2 k_{1} k_{2}+ n_2  k_{3} k_{2}+ n_1  k_{1}
   k_{3}}{n_1 k_{1}+n_2 k_{2}+ k_{3}} \Big).
\end{split} 
\ee
where $m_{n_1,n_2}=m_{n_1n_2+1}$.

So far we have used \eqref{final1}, \eqref{final2}. It is easy to show that the remaining 
equation \eqref{eq3spc} acting on 
the solution above gives rise to the two conditions 
\begin{eqnarray}
\sum_{n_1,n_2}m_{n_1,n_2}n_1^3 & = & 1, \label{condm1} \\
\sum_{n_1,n_2}m_{n_1,n_2} n_2^3 & = & 1. \label{condm2}
\end{eqnarray}

\subsection{Various Limits For The Momenta}
\label{variouslimits}
In this subsection we will show by considering two different limits for the momenta
that one can rule out
three of the four terms which appear in the sum in \eqref{ftform} leaving only the term with $n_1=n_2=1$.
The normalization of this term is then fixed by \eqref{condm1},\eqref{condm2} leading to  the  unique result 
given in  \eqref{valsm}.

\subsubsection{First  Limit } 

First consider the limit where the momentum carried by the tensor perturbation is much smaller than that of the
 two scalar perturbations,  
\be
\label{lim1aa}
k_3 \ll k_1 \simeq k_2. 
\ee
In this limit the scalar perturbations can be taken to be propagating in an essentially constant metric $\gamma_{ij}$.
The resulting wave function \eqref{wf3} can be calculated in two ways. Either by working 
directly with the  boundary values,  
$\gamma_{ij},\delta\phi$.
Or by first taking a  boundary metric which is flat, $\gamma_{ij}=\delta_{ij}$, and then transforming the
 answer
by a coordinate transformation to the case of the constant metric $\gamma_{ij}$. The two answers must
of course  agree.

This gives rise to the condition, \cite{Maldacena:2002vr},  that in this limit
\be
\label{condlim1a}
\langle T_{ij}(k_3) O(k_1) O(k_2)\rangle' e^{s,ij} =-e^{s,ij} k_{2i} k_{2j} {d\over dk_2^2} \langle O(k_2) O(-k_2)\rangle', 
\ee
where the superscript prime on the two sides stands for the correlator without the factor of $(2\pi)^3 \delta^3(\sum \vect{k_i})$. 
From \eqref{tpto}  and  \eqref{fsola} this gives that in the  limit \eqref{lim1aa}   
\be
\label{condlim1b}
S \rightarrow {3\over 2} k_2.
\ee
One finds that this condition rules out the two terms in  \eqref{ftform}
  where $n_1,n_2$  have the opposite sign so that  
\be
\label{formStwo}
\begin{split}
S = \sum_{\{(n_1, n_2)=(+,+), (n_1,n_2)=(-,-)\} } m_{n_1 n_2 } \Big(
 &-n_1 n_2  \frac{\norm{k_{2}} \norm{k_{3}}
   \norm{k_{1}}}{(n_1 \norm{k_{1}}+n_2 \norm{k_{2}}+\norm{k_{3}})^2}+n_1  \norm{k_{1}}+n_2 \norm{k_{2}}+
    \norm{k_{3}}\\
&-\frac{n_1 n_2 \norm{k_{1}} \norm{k_{2}}+ n_2  \norm{k_{3}} \norm{k_{2}}+ n_1  \norm{k_{1}}
   \norm{k_{3}}}{n_1 \norm{k_{1}}+n_2 \norm{k_{2}}+ \norm{k_{3}}} \Big).
\end{split}
\ee

\subsubsection{Second Limit  and the OPE}
\label{secondlimit}
Next we examine the limit where $k_2\simeq k_3 \gg k_1$. The  behaviour in this limit is most easily understood if 
we can appeal to the operator product expansion (OPE).  We have seen that the coefficient functions which appear in the 
wave function \eqref{wf1}, \eqref{wf2}, transform under the conformal symmetries like the correlation functions of a CFT. 
It is well known that in a CFT  operators satisfy the operator product expansion. For the  arguments that 
follow  we will 
 assume  that this is true for the coefficient functions in the wave function as well. While this assumption 
is quite plausibly true we do  not provide a proof for it here.\footnote{In the AdS/CFT correspondence which is related by analytic continuation to the dS case 
one can plausibly provide an argument for the operator product expansion from the bulk 
 using the prescription for calculating the boundary correlation functions from the bulk,
the properties of the bulk to boundary propagator, etc. By analytic continuation one would expect then to be 
able to show this for the 
coefficient functions  in the dS case as well.} In the next section we
provide another argument for uniqueness that does not reply on the OPE.

To see how the argument goes let us first examine the limit which was studied above, where $k_1,k_2$ are 
large compared to $k_3$, but now using the OPE.
We take
\be
\label{lim1large}
\vect{k}_2=\vect{K},  \ \ \vect{k}_1=-\vect{K}+\vect{k}_3, \ \  ~{\rm
  with}~ K \equiv |\vect{K}|\gg k_3.
\ee
 In position space we are considering the
 limit $\vect{x_1}\rightarrow \vect{x_2}$ for the correlation function 
\be
\label{corrfo}
\langle O(\vect{x_1}) O(\vect{x_2}) T_{\mu\nu}(\vect{x_3})\rangle.
\ee
 The operator product expansion tells us that in this limit the leading contribution comes from the term  
\be
\label{ope3}
O(0) O(\vect{x}) ={x_\mu x_\nu\over x^5} T^{\mu\nu}(\vect{x}) + \ldots,
\ee
where $\vect{x} \equiv \vect{x_2}-\vect{x_1}$.

The momentum space correlator is obtained by taking a Fourier
transform of \eqref{corrfo}
\be
\begin{split}
& \int \langle O(\vect{x_1}) O(\vect{x_2}) T^{\mu \nu}(\vect{x_3}) \rangle e^{i \left(\dotp[k_1, x_1] + \dotp[k_2, x_2] + \dotp[k_3, x_3] \right)} 
{d^3 x_1} {d^3 x_2} {d^3 x_3} \\ &= \int \langle O(0) O(\vect{x_2}-\vect{x_1})
T^{\mu \nu}(\vect{x_3}-\vect{x_1}) \rangle e^{i \left((\vect{k_1} + \vect{k_2} +
    \vect{k_3}) \cdot \vect{x_1} + \vect{k_2} \cdot  (\vect{x_2} - \vect{x_1})  + \vect{k_3} (\vect{x_3} - \vect{x_1}) \right)} {d^3 x_1} {d^3 x_2} {d^3 x_3} \\
&= (2 \pi)^3 \delta^3(\vect{k_1} + \vect{k_2} + \vect{k_3}) \int  \langle O(0) O(\vect{x_2}-\vect{x_1}) T^{\mu \nu}(\vect{x_3}-\vect{x_1}) \rangle 
e^{i \left( \vect{k_2} \cdot  (\vect{x_2} - \vect{x_1})  + \vect{k_3}
    \cdot  (\vect{x_3} - \vect{x_1}) \right)}  {d^3 x_2} {d^3 x_3}.
\end{split}
\ee

In the limit \eqref{lim1large} it follows from \eqref{ope3} that the momentum space correlator should go like 
\be
\int {x^{\mu} x^{\nu} \over x^5} e^{i \vect{K} \cdot \vect{x}} d^3 x \sim O(K^0).
\ee

Since the expression \eqref{fsola} already has a factor of $K^2$ outside, we learn that 
\be
\label{iopelargek1k2}
S \sim {k_3^3 \over K^2},
\ee
where we have inserted the correct factor of $k_3$ by dimensional analysis.  It is easy to check that this only 
happens in the sum in \eqref{ftform} if $n_1,  n_2$ have the same sign.  

For example, consider the term in \eqref{ftform} with $n_1 = n_2 = 1$. And scale $k_3 \rightarrow \lambda k_3$ and 
expand in powers of $\lambda$, for small $\lambda$.
We get:
\be
\label{ilargek1expansion}
\begin{split}
-S &= \frac{3 {k_{1}}}{2}+\frac{3 {\dotp[k_{1},k_{3}]} \lambda }{4
   {k_{1}}}+\left(\frac{{k_{3}}^2}{{k_{1}}}-\frac{{\dotp[k_1, k_3]}^2}{4 {k_{1}}^3}\right)
   \lambda ^2 \\ &+ \lambda^3 \left(\frac{3 {\dotp[k_1, k_3]}^3}{16 {k_1}^5}-\frac{9 {k_3}^2 {(\vect{k_1} \cdot \vect{k_3})}}{16
   {k_1}^3}-\frac{3 {k_3}^3}{8 {k_1}^2}\right).
\end{split}
\ee
One might naively believe that this contradicts \eqref{iopelargek1k2}. However, it is rather interesting that all the terms that grow too fast with $K$ are actually analytic {\em in at least two momenta} and so lead to contact terms when transformed to position space.

For example, we have
\be
\begin{split}
&\int k_{1 i} k_{2 j} {3 k_1  \over 2} \delta^3(\vect{k_1} + \vect{k_2} + \vect{k_3}) e^{-i \left(\vect{k_{1}} \cdot \vect{x_{1}} + \vect{k_{2}} \cdot \vect{x_{2}} + \vect{k_{3}} \cdot \vect{x_{3}} \right)}  d^3 k_1 d^3 k_2 d^3 k_3 \\ &=- {\partial \over \partial x_{1}^i} 
{\partial \over \partial x_2^j} \int {3 \norm{k_1} \over 2}  e^{-i \left(\vect{k_{1}} \cdot (\vect{x_{1}} - \vect{x_{2}})  + \vect{k_{3}} \cdot (\vect{x_{3}} - \vect{x_{2}}) \right)}  d^3 k_1 d^3 k_3 \\ &= - (2\pi)^3 
{\partial \over \partial x_{1}^i} {\partial \over \partial x_2^j} 
\delta(\vect{x}_3 - \vect{x}_2) \int {3 \norm{k_1} \over 2}  e^{-i
  \left(\vect{k_{1}} \cdot (\vect{x_{1}} - \vect{x_{2}}) \right)} d^3 k_1 + \ldots
\end{split}
\ee
where $\ldots$ are subleading in $\lambda$. 

The first non-analytic term in \eqref{ilargek1expansion} is the term that goes like ${k_3^3 \over k_1^2}$, which is indeed of the form that we expected in \eqref{iopelargek1k2}!

It is easy to check that if we consider a term in \eqref{ftform} where $n_1,n_2$ have opposite sign
we will not get an answer consistent with the OPE. For example consider the term  with $n_1 = -1, n_2 = n_3 = 1$, we have
\be
-S = {1 \over \lambda} \frac{2 {k_{3}} {k_{1}}^4+{\dotp[k_1, k_3]} {k_{1}}^3}{({\dotp[k_{1},k_{3}]}+{k_{1}}
   {k_{3}})^2}.
\ee
This is already non-analytic and is clearly of the wrong form.

Having considered the limit where $k_1,k_2$ are large compared to $k_3$ we can finally turn to the limit of interest where $k_2,k_3$ are large compared to $k_1$. In position space this corresponds to the 
 case where $x_2 \rightarrow x_3$, in which case
we expect the dominant OPE
\be
\label{ix2x3ope}
\begin{split}
O(x_2) T_{\mu \nu} (x_3) = &A {(x_2 - x_3)_{\mu} (x_2 - x_3)_{\nu} \over \norm{x_2 - x_3}^5} O(x_3)  
+ B {(x_2 - x_3)_{\mu} \partial_{\nu} +  (x_2 - x_3)_{\nu} \partial_\mu \over \norm{x_2 - x_3}^4} O(x_3) \\ &+ {C \over \norm{x_3}^3} \partial_{\mu} \partial_{\nu} O(x_3).
\end{split}
\ee

We are now concerned with the limit where $k_3 = K, k_2 = -K - k_1$ and ${\norm{K} \over \norm{k_1}}$ is large. The terms that multiply $A$ and $B$ might seem like they scale like $K^0$ in this limit, but this is deceptive. In fact, if we work through the Fourier transform, we expect that these terms give rise to
\be
{K_{\mu} K_{\nu} + K_{(\mu} k_{1 \nu)} + k_{1 \mu} k_{1 \nu} \over K^2},
\ee
in Fourier space. Of these terms only the last one --- $k_{1 \mu} k_{1 \nu}$ is meaningful, since the others point along $K$, and yield $0$ when contracted with a transverse polarization tensor for the stress-tensor. A similar logic applies for the term that multiplies $B$. So, in fact, all the three terms in \eqref{ix2x3ope} should give terms that scale like ${1 \over K^2}$ when transformed to momentum space.

This now implies that $S$ itself must scale like ${1 \over K^2}$
since  the full correlator is given by $S$ multiplied with 
$e^{s,i j} k_{1 i} k_{2 j}$ \eqref{fsola}, 
and even though $\vect{k_2}= -\vect{K} - \vect{k_1}$, since $e^{s,i j} K_j = 0$, 
this factor scales like $O(K^0)$.
It is now simple to see that of the two terms that remain in \eqref{formStwo} the only one which gives the correct behaviour for $S$ is the one with $n_1=n_2=1$. The analysis of expanding the terms in this limit and comparing with 
the required behaviour is completely analogous to the one above and we will skip the details. 

To summarize, by considering two limits for the momenta 
we learn that of the four terms which could have been present in $S$,
\eqref{ftform}  only one term survives giving the final result in \eqref{valsm}.

\subsection{Another Argument for Uniqueness}
Now, we give a second argument --- which does not assume the OPE --- for why only the choice $n_1 = 1, n_2 =
1$ is allowed. This argument closely follows an argument made in \cite{Raju:2012zr}.We will show that any correlator that arises from a
local interaction in de Sitter space, in the presence of the Bunch
Davies boundary conditions described in section \ref{symandcons} 
will have a pole in the quantity
$E = \norm{k_1} + \norm{k_2} + \norm{k_3}$. Demanding the existence of
this pole 
immediately tells us that the choice $n_1 = n_2 = 1$ is the only one
that is allowable.

The argument is as follows. The correlator that we are interested in is the coefficient of a particular term in the expansion of the wave-function of the Universe. First, let us consider the case where 
the correlator can be computed order by order in perturbation theory. Operatively, this means that we start with the solutions to the free equations of motion for the metric and the scalar perturbations, and then correct them perturbatively.\footnote{This is almost a universal approach to perturbation theory. For example, even in the Vasiliev theory, which involves an infinite number of derivatives, this is precisely how correlation functions are calculated. (See \cite{Giombi:2012he}, sections 4.2 and 4.3.)}

The solutions to the free equations of motion are given by:
\be
\label{wavefuncs}
\begin{split}
&\gamma_{i j} (\vect{k_3}) = \kappa_{\gamma} e^s_{i j} (1 - i \eta \norm{k_3}) e^{i
  \norm{k_3} \eta} e^{i \vect{k_3} \cdot \vect{x}}, \\
&\delta \phi (\vect{k_n}) =   \kappa_{\phi} (1 - i \eta \norm{k_n}) e^{i
  \norm{k_n} \eta} e^{i \vect{k_n} \cdot \vect{x}},
\end{split}
\ee
where $\kappa_{\gamma}$ and $\kappa_{\phi}$ are some constants and $\vect{k_n}$
may be either $\vect{k_1}$ or $\vect{k_2}$.  Note that,  as
we explained above, our choice of boundary conditions in the far past
fixes the sign of the exponent involving $\eta$.

 Without making {\em any} assumptions about the form of the interaction, the leading contribution to the correlator in such an expansion will be given by acting with some linear functional (which comes from the interaction vertices in  the action \eqref{action1}) on these perturbations. Denoting this linear functional by $S_{\text{int}}$, we have
\be
\label{varaction1}
\langle T^{i j}(\vect{k_3}) O(\vect{k_1}) O(\vect{k_2}) \rangle =
S_{\text{int}} [\gamma_{i j}(\vect{k_3}), \delta \phi(\vect{k_1}), \delta
\phi(\vect{k_2})].
\ee

Let us focus on the contribution to the correlator that comes from
{\em very early times} i.e. from $\eta \rightarrow -\infty$. While,
\eqref{action1} may be very complicated with many higher derivative
terms, in this limit the variation \eqref{varaction1} is controlled by
the highest power of $\eta$ that appears:
\be
\label{highesteta}
\langle T^{i j}(\vect{k_3}) O(\vect{k_1}) O(\vect{k_2}) \rangle = \int d \eta \, \eta^{m} e^{i \eta ( \norm{k_1} +
  \norm{k_2} + \norm{k_3})}  {\cal C}_f^{i j}(\vect{k_1}, \vect{k_2},
\vect{k_3}) + \ldots,
\ee
where the $\ldots$ indicate terms that come with lower powers of
$\eta$ and ${\cal C}^f$ is independent of $\eta$. We get a
contribution of $\eta^{-4}$ from the $\sqrt{-g}$ in the action. The $\eta^m$
arises by combining this term with the terms linear in $\eta$ in
\eqref{wavefuncs} and other factors of $\eta$ from the inverse
metric.  We can easily check that the two-derivative interaction
already gives rise to $m = 1$ and since higher derivatives require more factors of the
unperturbed inverse metric, they give rise to higher powers of
$\eta$. Evidently, doing
this integral leads to
\be
\langle T^{i j}(\vect{k_3}) O(\vect{k_1}) O(\vect{k_2}) \rangle =
{\Gamma[m+1] \over E^{m+1}} {\cal C}^{i j}_f(\vect{k_1}, \vect{k_2},
\vect{k_3}) + \ldots,
\ee
where $\ldots$ are now terms that have lower order poles in $E$.

With only a small amount of additional work, we can actually show that ${\cal C}^{i j}_f$ is related to the {\em flat space} scattering amplitude for two
scalars and a graviton. The reader can already
see this from the answer written in the form
\eqref{spinorhelans}. We refer the reader
to \cite{Raju:2012zr} (see section 5) for details on why we should have expected this. 

Returning to the problem at hand we see that since it is only $n_1 = n_2 = 1$ that gives the correct
pole, this is the only allowed choice. This proves the uniqueness of our
solution.

At this point, the reader may wonder why higher
  derivative terms, which would have given rise to higher poles, do not
  contribute in this answer. This is related to the fact that the
 form of the  three-point on-shell amplitude for two scalars and a
 graviton in four flat dimensions is exact, already at tree-level, and is not altered either  by higher-derivatives or loop corrections. 

It is quite simple to see this, since there is a unique Lorentz invariant that can be formed from the physical quantities at hand --- the polarization tensor of the graviton, and the three four-dimensional momenta $\vect{k_1}, \vect{k_2}, \vect{k_3}$. The on-shell condition tells us that $\vect{k_1}^2 = \vect{k_2}^2 = \vect{k_3}^2 = 0$ and since $\vect{k_1} + \vect{k_2} + \vect{k_3} = 0$, all dot products between the momenta vanish as well.\footnote{The momenta themselves do not have to be collinear if we allow them to be complex.} So, the only four dimensional Lorentz invariant we can form is $e^s_{i j} k_1^i k_2^j$. All that loop corrections or higher derivatives could possibly do is to renormalize the coefficient of this quantity but this is also fixed by the Ward identities. 

It is this fact about flat-space scattering amplitudes that is related to the uniqueness of our
 correlation function.

\subsection{Final Solution}
As mentioned above the unique solution for $S$ was obtained above in \eqref{valsm}.
The overall normalization followed from the use of the normalization of the two point function $\langle O(k_1) O(k_2)\rangle$
given in \eqref{tptO} which in turn determined the Ward identity \eqref{Ward}.

Instead  as discussed in  Appendix \ref{appendix1} it is convenient  to take   
 the two point function $\langle O(k_1) O(k_2)\rangle$ to be normalized as given 
in  \eqref{tpto} so that its normalization differs
 from \eqref{tptO}  by a factor of\footnote{The constant $c$ can be set to unity by rescaling the inflaton, but keeping it explicit allows for the normalization of the inflaton to be determined in an independent manner.} $c$. With this choice
    the solution for the correlator becomes 
\be
\label{fsolb}
\langle O(k_1) O(k_2) T_{ij}(k_3)\rangle e^{s, ij}=-2 (2\pi)^3 c \delta(\sum_i\vect{k}_i)e^{s, ij} k_{1i}k_{2j} S.
\ee

From the general arguments of section \ref{basicsetup} this should be the value for the 
coefficient function, $\langle O O T_{ij}\rangle e^{s, ij}$,  in the wave function \eqref{wf1}.

\section{Final Result}
Using the wave function \eqref{wf1}  and \eqref{infper}
 it is now  a simple matter to find the
 three point correlator involving two scalar perturbations $\zeta(\vect{k}_1), \zeta(\vect{k}_2)$ 
and one tensor perturbation  $\gamma_{ij}(\vect{k}_3)$ with polarization $e^{s, ij}$.

One finds that it  is given by 
\be
\label{fsolc}
\langle\zeta(k_1) \zeta(k_2) \gamma_s(k_3)\rangle = 
(2\pi)^3 \delta(\sum_i\vect{k}_i) {1\over \Pi_i(2k_i^3)}
({4H^4\over M_{pl}^4 c })({H^2 \over \dot{\phi}^2})
e^{s,ij} k_{1i}k_{2j}  S(k_1,k_2,k_3),
\ee
with 
\be
\label{fS}
S(k_1,k_2,k_3)=(k_1+k_2+k_3) - {\sum_{i>j}k_ik_j \over (k_1+k_2+k_3)} - {k_1k_2k_3\over (k_1+k_2+k_3)^2}.
\ee
In this formula $\dot{\phi}$ is the time derivative of the inflaton and  
 $c$ is  a constant which 
is defined from the normalization of the scalar two-point function given in \eqref{tpto}.
This constant  can be set to unity by rescaling $\dot{\phi}$. When the two derivative approximation is valid,  in the normalization 
where $c=1$,
$\dot{\phi}$ is related to the slow roll parameter $\epsilon$  by \eqref{relsla2}. 
$\gamma_s$ is related to the tensor  perturbation by 
\be
\label{relgs}
\gamma_{ij}(k_3) =\gamma_s(k_3)  e^s_{ij}(k_3),
\ee
where $e^s_{ij}(k_3)$ is the polarization which is  transverse and  traceless, 
\eqref{tratra},
with normalization given in \eqref{polnorm}. 

Equation \eqref{fsolc} is the main result of this paper.
 
By comparing this result   with the two point functions for the scalar and tensor perturbations given in
\eqref{tpto}, \eqref{tpstress} of the appendix \ref{appendix1}
  we see that the normalization of the correlator is completely
fixed in terms of the
normalization of these two two-point functions.

For conventional slow-roll inflation  the answer above agrees, up to an overall  sign,  with that obtained in 
\cite{Maldacena:2002vr}, with $c=1$ and $\phi$ being the canonically normalized inflaton.

\section{Conclusions}
In this paper we have  studied the three point function involving two scalars and one tensor perturbation.
We showed that this correlator is completely fixed by the $SO(1,4)$ symmetries of de Sitter space, up to small corrections. Our final result is given in \eqref{fsolc}. The normalizations for the scalar and tensor two point
 functions are given in \eqref{tpto} and \eqref{tpstress};  we see that  the normalization of the three point function is fixed in terms of the normalization of the two point functions. 

Our result is based on three  main  assumptions.
 First, that the inflationary dynamics--- including the scalar sector---approximately preserves
the full $SO(1,4)$ conformal group of isometries of de Sitter space. 
 Second, that 
there is only one scalar field during inflation. And third, that the initial state is the Bunch-Davies vacuum. 
Other than these assumptions the  result is general and  essentially model independent. 
In particular it  should apply to models where higher derivative corrections in gravity are important,
as  was discussed in the introduction. 

The general nature of this result means that    this three point function is observationally 
 a good way to test if the inflationary dynamics  had the full conformal group including the special conformal transformations as its
 symmetries.     
It is worth emphasizing that the two point functions do not 
by themselves  allow for a  test of  this feature. In conventional slow-roll inflation there is one relation between
the various parameters which arises as follows. The  
tensor two point function  allows for a determination of $H^2/M_{Pl}^2$ from its normalization and for  
$\epsilon$, defined in \eqref{srp}, from its tilt.  The normalization of the scalar two-point function goes like 
${H^2\over M_{Pl}^2} {H^2 \over \dot{\phi}^2}$ and is then 
 fixed since ${\dot{\phi}\over H}$ is determined in terms of $\epsilon$ by \eqref{relsla2}. 
However, once higher derivative corrections are included \eqref{relsla2}
 need not be valid any longer even when the full
conformal group is approximately preserved.  For example \eqref{relhv} could 
receive corrections due to higher
powers of curvature becoming important in the action \eqref{action1}. Thus this   relation between the parameters of the two point 
functions does not  allow us to test whether the special conformal transformations  were good symmetries during inflation.

Corrections to our result for the three point function will arise from effects which break the $SO(1,4)$ symmetries. 
These can be of two kinds.  
Effects which  break the special conformal symmetries but preserve scale invariance
and effects which break scale invariance.  Examples of the breaking of special conformal invariance include a speed of sound which is different from unity. More generally,  these effects can be parameterized
 using the effective Lagrangian approach discussed in \cite{Cheung:2007st}.
 The breaking of  scale invariance
occurs because the  Hubble constant and the  inflaton slowly evolve during inflation and are not 
constant. 
When the  momenta of the three perturbations in the correlator are of the same order of 
 magnitude one immediate way to incorporate some of the 
resulting corrections is to set the parameters, $H, \dot{\phi}$  which enter in \eqref{fsolc},
 to take their values
at the time of horizon crossing for the modes.\footnote{In the squeezed limit, when one  momentum is 
much smaller one can also incorporate similar effects by carrying out an analysis along  the lines of section 
\ref{variouslimits}.}
More generally, corrections  due to the breaking of  scale invariance 
 are of order the
slow roll parameters and  about $1\%$ in order of magnitude. 

As stated above our result applies to   models of single field inflation. When more than one scalar is 
present  both adiabatic
and isocurvature perturbations can be present  and it is harder to come up with model independent results. 
We can always still go to the gauge where $\zeta=0$, discussed for the single scalar case in \ref{gauge2}.
 And then work in a basis where the scalar
 field perturbations, $\delta \phi_i, i =1, \cdots N,$  have a diagonal two point function. Assuming that 
scalars are approximately massless we get the two-point functions to be\footnote{Here we have rescaled 
$\delta \phi$ to set a possible constant $c$ which appears in the normalization on the RHS to unity.} 
\be
\label{twomulti}
\langle\delta \phi_i(\vect{k}_1) \delta \phi_j(\vect{k}_2)\rangle=
\delta_{ij} (2 \pi)^3 \delta^3(\vect{k_1}+\vect{k_2}) {H^2 \over M_{pl}^2} {1\over 2 } {1\over k_1^3}.
\ee
  The three point functions for the scalar and tensor perturbations   then easily follows 
and is again diagonal in this basis of scalar perturbations and takes the model independent
form \eqref{fsolc}  (with $c=1$). 
The model dependence in the result enters when we try to obtain the three point function in terms of the 
the curvature perturbation, $\zeta$, 
 which is defined for all time and  conserved after the modes cross the horizon. 
 The value of this perturbation and its correlations depend on how the various scalars affect the 
end of inflation and this is    model dependent.

The analysis in this paper
 is based on earlier papers \cite{Maldacena:2002vr,Maldacena:2011nz}. In \cite{Maldacena:2011nz} it was shown that 
working in the de Sitter approximation the three point tensor perturbation can be significantly constrained from symmetry considerations alone.
Unlike tensor perturbations when dealing with scalars the small
breaking of de Sitter symmetries in the inflationary
background cannot be totally ignored. However for the correlation function of interest in this paper
 this breaking can be incorporated, at least  to leading order in the slow-roll parameters,
 in a straightforward manner. As explained in section \ref{basicsetup} one  first works in the  gauge 
 where $\zeta=0$ and calculates the correlation function in terms 
of the scalar perturbation $\delta \phi$, then transforms to the gauge where the 
 $\zeta \ne 0$ 
using \eqref{infper}. The calculation in terms of $\delta \phi$ can be done in de Sitter space and the 
breaking of de Sitter invariance enters only in the last step through the 
factor of ${\dot{\phi} \over H}$ in \eqref{infper}.  This is 
analogous to using the relation \eqref{tracerel}
 in conformal perturbation theory and computing the correlation function in terms of the scalar 
 operator $O$ in the CFT. 

It is  important to try  to extend this analysis to other correlation functions especially 
the three point scalar 
correlator which is observationally most significant. 
The analysis is more complicated here  since in general 
one cannot get away by simply taking the breaking of the de Sitter symmetries  into account in the  
manner described in the previous paragraph. 
This can be seen  from the results for the   conventional slow-roll case in  \cite{Maldacena:2002vr} where it was found  that the 
scalar three-point function is suppressed by an additional factor of $\sqrt{\epsilon}$ leading to an 
answer that goes like\footnote{This fact also follows from  CFT by noting that 
the three-point function of an exactly marginal operator must vanish.}   
 ${H^4 \over M_{Pl}^4 \epsilon}$. 
Despite these complications,  it would be worthwhile to consider a CFT which has  say just the stress 
tensor and a scalar as its low dimension
operators and ask how much the scalar  correlators are constrained by  CFT considerations alone 
along the lines of \cite{Heemskerk:2009pn}.

We have used both scale and special conformal invariance in deriving our result. 
  We have already discussed the possibility that   the scalar sector could break the special 
conformal symmetries badly. On the gravity side 
 translations,  rotations and   scale invariance uniquely lead 
to de Sitter space, which is then also   invariant under  special conformal transformations.
 However, more generally, when higher spin fields are also  excited it is conceivable that one has 
time dependent solutions   with translations, 
rotations and scale invariance symmetry but without  special conformal invariance. 
It would be worth developing an understanding of such solutions and their possible role in the 
early Universe.\footnote{For a review of higher spin fields and related issues see \cite{Vasiliev:2004cp}.}
 The correlator studied here   
could be used to distinguish solutions of this type also    from de Sitter space.

\acknowledgments
SPT thanks Yoskie Sumitomo for early collaboration. 
We are grateful to  N. Iizuka, E. Komatsu, S. Kachru,  J. Maldacena, G. Mandal, S. Minwalla, T. Souradeep
 and A. Yadav for discussions. IM thanks the organizers of the ICTS school on cosmology and gravity waves,
Dec. 2011.
SR and SPT thank the organizers of the ``strings discussion meeting''
at ICTS (Bangalore) for a 
stimulating conference where some of this research was carried out. SR
acknowledges the support of a Ramanujan Fellowship from the Department
of Science and Technology, Govt. of India. SR is grateful to Brown
University, where part of this work was carried out, for its hospitality.
SPT acknowledges the support of a J.C. Bose Fellowship from the Department. of Science and Technology,
Govt. of India.  
Most of all, we would all like to acknowledge our extensive debt to the
people of India.

\appendix
\section{The Two Point Function and Normalizations}
\label{appendix1}
In this appendix we discuss the two point function and related issues about normalizations of correlation functions. 
The wave function at quadratic order can be read off from \eqref{wf2} 
\be
\begin{split}
\psi  =  \exp\Bigg({M_{pl}^2\over H^2}\Big[&-{1\over 2}\int {d^3k \over (2\pi)^3}  {d^3k'\over (2\pi)^3}  
\delta\phi(\vect{k}) \delta\phi(\vect{k'}) 
\langle O(-\vect{k}) O(-\vect{k'})\rangle \\ 
&-{1\over 2} \int{ d^3k\over (2\pi)^3} {d^3k' \over (2\pi)^3}
\gamma_s(\vect{k})  \gamma_{s'}(\vect{k'}) \langle T^s(-\vect{k})
T^{s'}(-\vect{k'})\rangle\Big]\Bigg). \label{wf5}
\end{split}
\ee
Here the labels $s,s'$ denote  the two  polarizations of the graviton. 
In our notation a graviton can be written as a linear combination of its two polarizations
\be
\label{twopol}
\gamma_{ij}(\vect{k})=\sum_{s=1,2} \gamma_s e^s_{ij}(\vect{k}),
\ee
where the polarization tensors are normalized so that 
\be
\label{polnorm}
e^{s,ij}e^{s'}_{ij}=2\delta^{s,s'}.
\ee
For the stress energy tensor we  define
\be
\label{notpol}
T^s(\vect{k}) \equiv T_{ij}(\vect{k})e^{s,ij}(-\vect{k}).
\ee

Translational and rotational invariance along with scaling symmetry fixes the form of the two point functions to be 
\begin{eqnarray}
\langle O(\vect{k_1}) O(\vect{k_2})\rangle&=&  c  \norm{k_1}^3 (2\pi)^3\delta^3(\vect{k_1}+\vect{k_2}), \label{tpto} \\
\langle T^s(\vect{k_1}) T^{s'}(\vect{k_2})\rangle& = &   \norm{k_1}^3 (2\pi)^3  \delta^3(\vect{k_1}+\vect{k_2}) ({ \delta^{s s'} \over 2} ). \label{tpstress}
\end{eqnarray}
A constant could  have appeared on the RHS of \eqref{tpstress} but that can be absorbed into a
 redefinition of $H$. The constant $c$ which appears on the RHS of \eqref{tpto} could also have been
 set to unity by rescaling the operator $O$. However doing so  also requires us to rescale the 
inflaton perturbation $\delta \phi$ which is the source for $O$.  It is   convenient instead to 
 not do this rescaling and keep the constant $c$ explicit  in \eqref{tpto}.

Substituting \eqref{tpto}, \eqref{tpstress} in the wave function one can easily show that the resulting two-point functions
for the perturbations are 
\begin{eqnarray}
\langle\delta \phi(\vect{k_1}) \delta \phi(\vect{k_2})\rangle & = & 
(2 \pi)^3 \delta^3(\vect{k_1}+\vect{k_2}) {H^2 \over M_{pl}^2} {1\over 2 c} {1\over k_1^3}, \label{tptphi} \\
\langle\gamma_s(\vect{k_1}) \gamma_{s'}(\vect{k_2})\rangle & = & 
(2\pi)^3 \delta(\vect{k_1}+\vect{k_2}) {H^2 \over M_{pl}^2}  {1\over 2 k_1^3} (2 \delta_{s,s'}). \label{tptgamma}
\end{eqnarray}
Using \eqref{infper}  we get from \eqref{tptphi} that the two point function of the scalar perturbation is 
\be
\label{tptzeta}
\langle\zeta(\vect{k_1})\zeta(\vect{k_2})\rangle=(2\pi)^3\delta^3(\vect{k_1}+\vect{k_2}) { H^2 \over M_{pl}^2} {1\over  2c} 
{H^2 \over \dot{\phi}^2} {1\over k_1^3}.
\ee
\eqref{tptgamma}, \eqref{tptzeta} agree with the results of the standard slow-roll 
 two -derivative theory when $c=1$
and $\phi$ is the canonically normalized inflaton. 
More generally $c$ can be set to unity by rescaling $\phi$. 

\section{Details of  the Equations for Special Conformal Invariance}
\label{appendix2}

From \eqref{formcorr} and \eqref{defM}  we learn that 
\begin{eqnarray}
M_{ij}(\vect{k}_1,\vect{k}_2,\vect{k}_3)& = & k_{1i} k_{1j} f_1(k_1,k_2,k_3)+
k_{2i}k_{2j} f_1(k_2,k_1,k_3) \nonumber  \\
 && +(k_{1i}k_{2j} + k_{2i} k_{1j}) f_2(k_1,k_2,k_3)
 + \delta_{ij} f_3(k_1,k_2,k_3).  
\end{eqnarray}
Multiplying by $ k_{3i}(k_{1j}-\frac{k_{3j}(k_1.k_3)}{k_3^2}) $ we get
\begin{equation}
\label{iap2}
k_{3i}(k_{1j}-\frac{k_{3j}(k_1.k_3)}{k_3^2})M_{ij}(\vect{k}_1,\vect{k}_2,\vect{k}
_3)=[k_1^2-\frac{\dotp[k_1, k_3]^2}{k_3^2}]\left(\dotp[k_1, k_3](f_1-f_2) + \dotp[k_2,k_3](f_2-f_1^T)\right).
\end{equation}

Now, choosing $\vect{b} \propto \vect{k}_1 - \frac{\vect{k_3}(\vect{k}_1\cdot \vect{k}_3)}{k_3^2} $ in \eqref{finalspc}
  we get 
\be
\label{apishan1}
\begin{split}
4 k_{1i} k_{1j} e^{s,ij} \Big[&(1+k_1\ddk ) f_1 -(1+k_2 \ddkk ) f_1^T
+ (k_2 \ddkk -k_1 \ddk) f_2   \\
 &-\frac{\dotp[\vk, \vkkk]}{k_3} \ddkkk (f_1 - f_2) +
 \frac{\dotp[\vkkk,  \vkk]}{k_3} \ddkkk (f_1^T-f_2) \Big]   \\ + k_{1i} k_{2j} 
e^{s,ij}\Big[&k_1^2-\frac{\dotp[\vk,  \vkkk]}{k_3^2}\Big](\Theta (k_1) - \Theta
(k_2) ) (2f_2 - f_1 -f_1^T) =0.
\end{split}
\ee
Using $ k_{2j} e^{s,ij} = -(k_{1j} +k_{3j} ) e^{s,ij} = -k_{1j} e^{s,ij}$, \eqref{apishan1}
  reduces to
\be
\label{iap7}  
\begin{split}
e^{s,ij} k_{1i} k_{1j} \Bigg\{&4[(1+k_1 \ddk)f_1 - (1+ k_2 \ddkk) f_1^T 
+ (k_2\ddkk -k_1\ddk) f_2 \\ &-\frac{\dotp[\vk, \vkkk]}{k_3} \ddkkk (f_1 - f_2 ) 
+ \frac{\dotp[\vkk, \vkkk]}{k_3} \ddkkk (f_1^T-f_2)] \\ &- (k_1^2 -\frac{\dotp[\vk, \vkkk]^2}{k_3^2})
(\Theta (k_1) - \Theta (k_2) ) (2f_2 - f_1 -f_1^T) \Bigg\}=0.
\end{split}
\ee
Since the polarization  can be chosen so that $e^{s,ij}k_{1i}k_{1j}$
does not vanish the quantity within the curly brackets must vanish
leading to
\be
\label{iap33}
\begin{split}
4\Big[&(k_1 \ddk-\frac{\dotp[\vk,  \vkkk]}{k_3} \ddkkk)(f_1-f_2)- (k_2 \ddkk-\frac{\dotp[\vkk,\vkkk]}{k_3} 
\ddkkk)(f_1^T-f_2)  \\
 &+(f_1-f_2)-(f_1^T-f_2) \Big] 
- (k_1^2 -\frac{\dotp[\vk, \vkkk]^2}{k_3^2})(\Theta (k_1) - \Theta (k_2) ) 
(2f_2 - f_1 -f_1^T) = 0.
\end{split}
\ee
In terms of $S\equiv [(f_1-f_2)+(f_1^T-f_2)]/2 $ and $A\equiv
[(f_1-f_2)-(f_1^T-f_2)]/2 $ this becomes
\be
\label{iap34}
\begin{split}
& 4\Big[(k_1 \ddk-\frac{\vk \cdot \vkkk}{k_3} \ddkkk)(S+A)- (k_2 \ddkk-\frac{\vkk \cdot \vkkk}{k_3} 
\ddkkk)(S-A)+2A \Big] \\
&- 2 (k_1^2 -\frac{(\vk \cdot \vkkk)^2}{k_3^2})(\Theta (k_1) - \Theta
(k_2) )  S  =  0.
\end{split}
\ee
Similarly, \eqref{iap2}  in terms of $S,A$ becomes,
\be
\label{iap35}
k_{3i}(k_{1j}-\frac{k_{3j}\dotp[\vk, \vkkk]}{k_3^2})M_{ij}(\vect{k}_1,\vect{k}_2,\vect{k}
_3)=[k_1^2-\frac{\dotp[k_1, k_3]^2}{k_3^2}](\dotp[k_1, k_3](S+A)   - \dotp[k_2,k_3](S-A)),
\ee
which can be used to solve for $A$ and gives 
\begin{equation}
A=\frac{(\vect{\vk}-\vect{\vkk})\cdot \vect{k_3}}{k_3^2} S
-\frac{M_{ij}k_{3i} \ep_{\perp j}}
{k_3^2 \ep_{\perp}^2},
 \end{equation} 
where $ \ep_{\perp j} \equiv
k_{1j}-\frac{k_{3j}\dotp[\vk,\vkkk]}{k_3^2}$.(We caution the reader
that this is different from the null transverse vector $\ep_3$ that
has appeared above.)
Substituting in \eqref{iap34} this leads to 
\be
\label{iapp9}
\begin{split}
4\Big[&(\frac{\vkk \cdot \vkkk}{k_3^2} k_1 \ddk - \frac{\vk \cdot \vkkk}{k_3^2} k_2 \ddkk) S 
+ \frac{\vkkk \cdot (\vkk-\vk)}{k_3^2}S \\ &+ (k_1 \ddk + k_2 \ddkk + k_3 \ddkkk) 
(\frac{M_{ij}k_{3i} \ep_{\perp j}}{2k_3^2 \ep_{\perp}^2})+  \frac{M_{ij}k_{3i}\ep_{\perp j}}{k_3^2 \ep_{\perp}^2}\Big] 
\\ &+ (k_1^2 -\frac{(\vk \cdot \vkkk)^2}{k_3^2})(\Theta (k_1) - \Theta (k_2) ) S 
=0. 
\end{split}
\ee

Next, using the Ward identity, \eqref{Ward} we get 
 \begin{equation}
\label{iap10}
 \frac{M_{ij}k_{3i}\ep_{\perp j}}{ \ep_{\perp}^2} = -k_1^3 + k_2^3.
\end{equation}
Substituting \eqref{iap10} in \eqref{iapp9} after some algebra gives \eqref{eq2spc}. 

Finally we consider taking $\vect{b}$ to be orthogonal to all
$\vect{k}_i$ so that $\vect{b} \cdot \vect{k}=0$. We can also choose a
polarization so that $b_i k_{1j} e^{s,ij}\ne 0$. \eqref{finalspc} then gives 
\be
\label{iap11}
\begin{split}
&(1+ k_1\partial_{k_1}) f_1 -(1+k_2 \partial_{k_2}) f_1^T    
+(k_2 \partial_{k_2}-k_1 \partial_{k_1})f_2 - {\dotp[k_3, k_1] \over k_3} \partial_{k_3}(f_1-f_2) 
\\ &+ {\dotp[k_3, k_2] \over k_3} \partial_{k_3}(f_1^T-f_2)=  0.
\end{split}
\ee
The reader will notice that the LHS above is the first two lines of the LHS of \eqref{iap7}. Thus the analysis above when applied to \eqref{iap11} directly leads  
to \eqref{eq3spc}.

\section{Spinor Helicity Formalism \label{appspinor}}
In this section we provide some further details on the calculations
of  subsection \ref{secspinhelconf}.  We will use the spinor helicity
formalism  that was first introduced in  \cite{Maldacena:2011nz}
although our notation is similar to that of \cite{Raju:2012zs}. The
paper \cite{Raju:2012zs} analyzed conformal field theory correlators
in a Lorentzian spacetime; here our correlators obey the constraints
of conformal invariance in a Euclidean spacetime. However, most of
the formalism carries over directly as we show below.

\subsection{Notation}
Given a Euclidean 3-momentum $\vect{k} = (k_1, k_2, k_3)$, we convert it into 
spinors using
\begin{equation}
k_{\alpha \dot{\alpha}} =\norm{k} \sigma^0_{\alpha \dot{\alpha}} +  k_1
\sigma^1_{\alpha \dot{\alpha}} +  k_2 \sigma^2_{\alpha \dot{\alpha}} +
k_3 \sigma^3_{\alpha \dot{\alpha}} = \la_{\alpha} \lb_{\dot{\alpha}},
\end{equation}
where
\begin{equation}
|\vect{k}| \equiv \sqrt{\vect{k} \cdot \vect{k}} = \sqrt{k_1^2 + k_2^2 + k_3^2}.
\end{equation}

We can raise and lower spinor indices using the $\epsilon$ tensor. We choose the $\epsilon$
tensor to be $i \sigma_2$ for both the dotted and the undotted indices. This means that 
\begin{equation}
\epsilon^{0 1} = 1 = -\epsilon^{1 0},
\end{equation}
and spinor dot products are defined via 
\be
\dotl[\la_1, \la_2] = \epsilon^{\alpha \beta} \la_{1 \alpha} \la_{2 \beta} = \la_{1 \alpha} \la_{2}^{\alpha}, \quad \dotl[\lb_1, \lb_2] = \epsilon^{\dot{\alpha} \dot{\beta}} \lb_{1 \dot{\alpha}} \lb_{2 \dot{\beta}} = \lb_{1 \dot{\alpha}} \lb_2^{\dot{\alpha}}. 
\ee

In the case of four-dimensional flat-space scattering amplitudes, all
expressions can be written in terms of the two kinds of dot products
above. However, in our case, we should expect our expressions for CFT$_3$ correlators to only have a manifest $SO(3)$ invariance. This means that we might have mixed products between dotted and undotted indices. Such a mixed product extracts the $z$-component
of vector and is performed by
contracting with $\sigma^0$
\begin{equation}
\label{mixedproduct}
2  |\vect{k}| = (\sigma^0)^{\alpha \dot{\alpha}} k_{\alpha \dot{\alpha}} \equiv  \dotlm[\la, \lb].
\end{equation}
The reader should note that we use square brackets only for this
mixed product; products of both left and right handed spinors are denoted by angular brackets. Second, we note that this mixed dot product is symmetric:
\begin{equation}
\dotlm[\la, \lb] = \dotlm[\lb, \la].
\end{equation}

When we take the dot products of two 3-momenta, we have
\begin{equation}
\label{dotproductspinor}
\begin{split}
&\vect{k} \cdot \vect{q} \equiv \bigl(k_1 q_1 + k_2 q_2 + k_3 q_3 \bigr) \\ &=  -{1 \over 2} \Big(\dotl[\la_k, \la_q] \dotlb[\lb_k, \lb_q] - {1\over 2} \dotlm[\la_k,\lb_k] \dotlm[\la_q,\lb_q] \Big).
\end{split}
\end{equation}

Another fact to keep in mind is that
\begin{equation}
\begin{split}
&\vect{k_1} + \vect{k_2} = \vect{k_3} \\ &\Rightarrow \la_1 \lb_1 + \la_2 \lb_2 = \la_3 \lb_3 + {1 \over 2} \bigl(\dotlm[\la_1,\lb_1] +  \dotlm[\la_2,\lb_2] - \dotlm[\la_3,\lb_3] \bigr) \sigma^0.
\end{split}
\end{equation}

We also need a way to convert dotted to undotted indices.  We write
\begin{equation}
\lad_{\dot{\alpha}} = -\sigma^0_{\alpha \dot{\alpha}} \la^{\alpha}, \quad \lbd_{\alpha} = -\sigma^0_{\alpha \dot{\alpha}} \lb^{\dot{\alpha}}.
\end{equation}
This has the property that
\begin{equation}
\dotlb[\mb, \lad] = \dotlm[\mb,\la],
\end{equation}
where the quantity on the right hand side is defined in \eqref{mixedproduct}.

With all this, we can write down polarization vectors for conserved currents. The polarization vectors for a momentum vector $\vect{k}$ associated with
spinors $\la, \lb$ are given by
\begin{equation}
\label{polarizationvects2}
\begin{split}
&\epsilon^+_{\alpha \dot{\alpha}} = 2 {\lbd_{\alpha} \lb_{\dot{\alpha}} \over \dotlm[ \la, \lb]} =  {\lbd_{\alpha} \lb_{\dot{\alpha}} \over  \norm{k}}, \\ 
&\epsilon^-_{\alpha \dot{\alpha}} =  2 {\la_{\alpha} \lad_{\dot{\alpha}} \over \dotlm[\la, \lb]} = {\la_{\alpha} \lad_{\dot{\alpha}} \over  \norm{k}}.
\end{split}
\end{equation}
These vectors are normalized so that 
\begin{equation}
\label{normalizationpol}
\vect{\epsilon^+} \cdot \vect{\epsilon^{+}} = \vect{\epsilon^-} \cdot \vect{\epsilon^{-}} = 0, \quad \vect{\epsilon^+} \cdot \vect{\epsilon^{-}} = 2.
\end{equation}
Polarization tensors for the stress tensor are just outer-products of these vectors with themselves:
\be
\label{gravpolten}
e^{\pm}_{i j} = \ep^{\pm}_i \ep^{\pm}_j.
\ee
We again caution the reader that these are normalized {\em differently} from
the $e^s_{i j}$ tensors, which appeared previously. These $e^{\pm}_{i
  j}$ tensors are linear combinations of those that correspond to
``circularly polarized'' gravitons. 

\subsection{Conformal generators in momentum space}
As a prelude to understanding the action of conformal generators using spinor helicity variables, we need expressions for the conformal generators in momentum space. In position space we have \cite{DiFrancesco:1997nk}
\be
\begin{split}
D &= -i x^{i} { \partial \over \partial x^{i}} - i \Delta, \\
K_{i} &= -2 i x_{i} \Delta - x^{j} S_{i j} - 2 i x_{i} x^{j} \partial_{j} + i x^2 \partial_{i}.
\end{split}
\ee
When we Fourier transform this, we should replace $x^{i} \rightarrow i { \partial \over \partial k_{i}}$ and ${\partial \over \partial x^{i}} \rightarrow i k_{i}$. These replacements lead to
\be
\label{conformalmom}
\begin{split}
D &= i {\partial \over \partial k_{i}} k_{i} - i \Delta = i(d - \Delta) + i k_{i} {\partial \over \partial k_{i}} \\
K_{i} &= 2 \Delta {\partial \over \partial k^{i}} - i S_{i j} {\partial \over \partial k_{j}} - 2 {\partial \over \partial k^{i}} {\partial \over \partial k_{j}}  k_{j} + {\partial \over \partial k^{j}} {\partial \over \partial k_{j}} k_{i} \\ &=  2 \Delta {\partial \over \partial k^{i}} - i S_{i j} {\partial \over \partial k_{j}} - 2 k_{j}  {\partial \over \partial k^{i}} {\partial \over \partial k_{j}}  +  k_{i} {\partial \over \partial k^{j}} {\partial \over \partial k_{j}} - 2 (d + 1 - 1) {\partial \over \partial k^{i}}  \\ 
&=  2 (\Delta - d) {\partial \over \partial k^{i}} - i S_{i j} {\partial \over \partial k_{j}} - 2 k_{j}  {\partial \over \partial k^{i}} {\partial \over \partial k_{j}}  +  k_{i} {\partial \over \partial k^{j}} {\partial \over \partial k_{j}}.
\end{split}
\ee
The $D$ above should be distinguished from the $\tilde{D}$ in \eqref{trspt}. 
For scalars, this can be recast as:
\be
\label{confonscalars}
\begin{split}
K^s_{i} &=  2 (\Delta - d) {k_{i} \over \norm{k}} {\partial \over \partial \norm{k}} - 2 k_{j}  {\partial \over \partial k^{i}} {k^{j} \over \norm{k}}  {\partial \over \partial \norm{k}}  +  k_{i} {\partial \over \partial k^{j}}  {k^{j} \over \norm{k}}  {\partial \over \partial \norm{k}} \\
&=  2 (\Delta - d) {k_{i} \over \norm{k}} {\partial \over \partial \norm{k}} - 2 {k_{i} \over \norm{k}}  {\partial \over \partial \norm{k}} - 2 k_{i} \norm{k}   {\partial \over \partial \norm{k}} {1 \over \norm{k}}  {\partial \over \partial \norm{k}}  +  {d k_{i} \over  \norm{k}}  {\partial \over \partial \norm{k}} +  k_{i}   \norm{k} {\partial \over \partial \norm{k}} {1  \over \norm{k}}  {\partial \over \partial \norm{k}} \\
&=  \left(2 \Delta - d - 1\right) {k_{i} \over \norm{k}} {\partial \over \partial \norm{k}} - {k_{i}  {\partial^2 \over \partial \norm{k}^2}}.
\end{split}
\ee
Here, we have just systematically replaced momentum derivatives using
\be
\label{momtonorm}
{\partial \over \partial k_{i}} = {\partial \norm{k} \over \partial k_{i}} {\partial \over \partial \norm{k}} = {k^{i} \over \norm{k}}  {\partial \over \partial \norm{k}},
\ee
which is true for functions that depend only on $\norm{k}$. 
\subsection{Conformal generators in spinor helicity variables for scalars}
Now we analyze how the double derivative operator ${\partial
  \over \partial \la_{\alpha}} {\partial \over \partial
  \lb_{\dot{\alpha}}}$ can be used like the generator of special conformal transformations.
Consider the object
\be
\tilde{K}_{i} = 2 {\partial \over \partial \lambda_{\alpha}} {\partial \over \partial \bar{\lambda}_{\dot{\alpha}}} \sigma_{i \alpha \dot{\alpha}}.
\ee
We can convert these derivatives to momentum derivatives.
Recall that we have 
\be
\label{momspinors}
\la_{\beta} \lb_{\dot{\beta}} = k_{m} \sigma^{m}_{\beta \dot{\beta}}  + \norm{k} \sigma^{0}_{\beta \dot{\beta}},
\ee
and also,
\be
k_{j} = {1 \over 2} \la_{\alpha} \la_{\dot{\alpha}} \bar{\sigma}_j^{\dot{\alpha} \alpha}.
\ee
This allows us to convert the spinorial derivatives to momentum derivatives as follows.
\be
\begin{split}
&\sigma_{i \alpha \dot{\alpha}} {\partial \over \partial \lambda_{\alpha}} {\partial \over \partial \bar{\lambda}_{\dot{\alpha}}} = \sigma_{i \alpha \dot{\alpha}} {\partial \over \partial \lambda_{\alpha}}   {\partial k_{j} \over \partial \bar{\lambda}_{\dot{\alpha}}} {\partial \over \partial k_{j}} \\
&= \sigma_{i \alpha \dot{\alpha}} {\partial \over \partial \lambda_{\alpha}}   {1 \over 2} \bar{\sigma}_{j}^{\dot{\alpha} \beta} \la_{\beta} {\partial \over \partial k_{j}} =  
{1 \over 2} \sigma_{i \alpha \dot{\alpha}}  \left( \bar{\sigma}_{j}^{\dot{\alpha} \alpha}  {\partial \over \partial k_{j}} + 
  {1 \over 2} \bar{\sigma}_{j}^{\dot{\alpha} \beta} \la_{\beta}  \bar{\sigma}_{k}^{\dot{\beta} \alpha} \lb_{\dot{\beta}} {\partial \over \partial k_{k}}   {\partial \over \partial k_{j}} \right).
\end{split}
\ee
In this expression, it is {\em important that the spacetime indices on $\sigma$ are summed only over $(1,2,3)$ i.e they are not summed over the 0-direction.}

Now, we note that 
\be
\begin{split}
&\sigma_{i \alpha \dot{\alpha}} \bar{\sigma}_{j}^{\dot{\alpha} \alpha} = 2 \eta_{i j} \\
&\sigma_{i \alpha \dot{\alpha}}  \bar{\sigma}_{j}^{\dot{\alpha} \beta}   \sigma_{m \beta \dot{\beta}} \bar{\sigma}_{k}^{\dot{\beta} \alpha} = 2\left(\eta_{i j} \eta_{k m} + \eta_{i k} \eta_{j m} - \eta_{i m} \eta_{k j} + i \epsilon_{i j k m} \right).
\end{split}
\ee

The totally antisymmetric term is not important since our expression is symmetric in $j$ and $k$. Second, note that the term involving $\sigma^0$ in \eqref{momspinors} drops out since the expression above involves a trace over products of $\sigma$ matrices and none of the other $\sigma$ matrices take the value $\sigma^0$ and, as we have already noted, the $\ep$-tensor term is unimportant.

Using all this, we find that 
\be
\label{specialconformalspins}
\tilde{K}_{i} =   2 {\partial \over \partial k^{i}} + 
   2 k_{j} {\partial \over \partial k_{j}} {\partial \over \partial k^{i}} -  k_{i} {\partial \over \partial k_{j}} {\partial \over \partial k^{j}} .
\ee
We see that $\tilde{K}_i$ agrees with the form \eqref{conformalmom} where $\Delta = d - 1$. (Up to an overall minus sign.)

Using \eqref{confonscalars}, for $\Delta = 2$ and $d = 3$ 
this is the statement that
\be
2 b_i \sigma^i_{\alpha \dot{\alpha}} {\partial \over \partial \la_{\alpha}} {\partial \over \partial \lb_{\dot{\alpha}}}  = \dotp[b,k] {\partial \over \partial \norm{k}^2}.
\ee

Now, consider a {\em marginal scalar} --- $O$. (This has dimension $d$). We note that
\be
2 b_i \sigma^i_{\alpha \dot{\alpha}} {\partial \over \partial \la_{\alpha}} {\partial \over \partial \lb_{\dot{\alpha}}} {O \over \norm{k}}  = \dotp[b,k] {\partial \over \partial \norm{k}^2} {O \over \norm{k}} = \dotp[b,k] \left({1 \over \norm{k}} {\partial^2 O \over \partial \norm{k}^2} - {2 \over \norm{k}^2} {\partial O \over \partial \norm{k}} + {2 O \over \norm{k}^3} \right).
\ee
Comparing with \eqref{confonscalars}, this means that 
\be
\label{doubleonmarginal}
2 b_i \sigma^i_{\alpha \dot{\alpha}} {\partial \over \partial \la_{\alpha}} {\partial \over \partial \lb_{\dot{\alpha}}} {O \over \norm{k}} = {1 \over \norm{k}} (-\vect{b} \cdot \vect{K^s}) O  + (\vect{b} \cdot \vect{k}) {2 O \over \norm{k}^3}.
\ee

Equation \eqref{doubleonmarginal} tells us that if we act with the
double derivative on a marginal scalar, divided by the appropriate
power of $\norm{k}$, we will still get a term on the right hand
side. This is similar to the``Ward identity'' term we get below for stress tensors, except here we find that the right hand side is proportional to the original correlator itself.

\subsection{Special conformal transformations on stress tensors in spinor-helicity variables}
We would now like to determine how the double derivative acts on
tensors contracted with polarization vectors. 

Using \eqref{gravpolten}, we write the polarization tensor as:
\be
e^{-}_{i j} = {1 \over \norm{k}^2} \bar{\sigma}_i^{a \dot{a}} \bar{\sigma}_j^{b \dot{b}} \la_{a} \lad_{\dot{a}} \la_{b} \lad^{\dot{b}},
\ee
Here, as opposed to \cite{Raju:2012zs}, we are also being
careful to denote sigma matrices with indices raised with a bar. This
is simply a matter of convenience.

We would like to calculate:
\be
2 b_k \sigma^k_{\alpha \dot{\alpha}}{\partial \over \partial \la_{\alpha}} {\partial \over \partial \lb_{\dot{\alpha}}} {e^{-}_{i j} \over \norm{k}^p} T^{i j},
\ee
where $p$ is a power of the momentum that we will fix for convenience later.
It is convenient to define 
\be
\tilde{e}^{-}_{i j} \equiv \norm{k}^2 e^{-}_{i j}; \tilde{T}^{i j} \equiv {T^{i j} \over \norm{k}^{p+2}}
\ee
 and instead compute:
\be
\label{confontwithpol}
\begin{split}
&2 b_k \sigma^k_{\alpha \dot{\alpha}} {\partial \over \partial \la_{\alpha}} {\partial \over \partial \lb_{\dot{\alpha}}} {\tilde{e}^{-}_{i j}} {\tilde{T}^{i j}} 
 = 2 b_k \sigma^k_{\alpha \dot{\alpha}} \left({\tilde{e}^{-}_{i j} }  {\partial \over \partial \la_{\alpha}} {\partial \over \partial \lb_{\dot{\alpha}}} {\tilde{T}^{i j}} +  \left({\partial \over \partial \la_{\alpha}} {\tilde{e}^{-}_{i j} } \right) {\partial \over \partial \lb_{\dot{\alpha}}} {\tilde{T}^{i j}} \right).
\end{split}
\ee
Here we have used the fact that $\tilde{e}^{-}_{i j}$ has no dependence on $\lb$.
Let us parse the various terms in this expression. We have
\be
\label{polderexp}
\begin{split}
&{\partial \over \partial \la_{\alpha}} {\tilde{e}^{-}_{i j}} =  
\left( {\lad_{\dot{a}} \la_{b} \lad_{\dot{b}} \bar{\sigma}^{\alpha \dot{a}}_i \bar{\sigma}^{b \dot{b}}_j   - \la_{a} \sigma^0_{\dot{a} \beta} \epsilon^{\alpha \beta} \la_{b} \lad_{\dot{b}} \bar{\sigma}^{a \dot{a}}_i \bar{\sigma}^{b \dot{b}}_j } \right) + (i \leftrightarrow j),
\end{split}
\ee
and
\be
\label{spindertomom}
{\partial {\tilde{T}^{i j}} \over \partial \lb_{\dot{\alpha}}} = {1 \over 2} {\partial {\tilde{T}^{i j}} \over \partial k_{m}} \bar{\sigma}_{m}^{\beta \dot{\alpha}} \la_{\beta}. \\
\ee
When we put the two equations above together, we encounter the term
\be
\label{polder1}
 b_k \sigma^k_{\alpha \dot{\alpha}} {\lad_{\dot{a}} \la_{b} \lad_{\dot{b}} \bar{\sigma}^{\alpha \dot{a}}_i \bar{\sigma}^{b \dot{b}}_j }  {\partial \tilde{T}^{i j} \over \partial k_{m}} \bar{\sigma}_{m}^{\beta \dot{\alpha}} \la_{\beta} = b_m \tilde{e}^{-}_{i j} {\partial \tilde{T}^{i j}  \over \partial k_{m}} + b_i \tilde{e}^{-}_{m j} {\partial  \tilde{T}^{i j}  \over \partial k_m} - b_k \tilde{e}^{-}_{k j} {\partial \tilde{T}^{i j} \over \partial k^i},
\ee
where we have used the identity (See 2.43 in \cite{Dreiner:2008tw})
\be
\begin{split}
&\sigma^{\mu} \bar{\sigma}^{\nu} \sigma^{\rho} = \eta^{\mu \nu} \sigma^{\rho} - \eta^{\mu \rho} \sigma^{\nu} + \eta^{\nu \rho} \sigma^{\mu} + i \epsilon^{\mu \nu \rho \kappa} \sigma_{\kappa}, \\
&\bar{\sigma}^{\mu} \sigma^{\nu} \bar{\sigma}^{\rho} = \eta^{\mu \nu} \bar{\sigma}^{\rho} - \eta^{\mu \rho} \bar{\sigma}^{\nu} + \eta^{\nu \rho} \bar{\sigma}^{\mu} - i \epsilon^{\mu \nu \rho \kappa} \bar{\sigma}_{\kappa},
\end{split}
\ee
and noticed that the last $\epsilon$ does not contribute since all indices are summed only over three dimensions here.
With a few index gymnastics we can check that, also
\be
\label{polder2}
 -b_k \sigma^k_{\alpha \dot{\alpha}}   {\la_{a} \sigma^0_{\dot{a} \gamma} \epsilon^{\alpha \gamma} \la_{b} \lad_{\dot{b}} \bar{\sigma}^{a \dot{a}}_i \bar{\sigma}^{b \dot{b}}_j } {\partial \tilde{T}^{i j} \over \partial k_{m}} \bar{\sigma}_{m}^{\beta \dot{\alpha}} \la_{\beta}= b_m \tilde{e}^{-}_{i j} {\partial \tilde{T}^{i j}  \over \partial k_{m}} + b_i \tilde{e}^{-}_{m j} {\partial  \tilde{T}^{i j}  \over \partial k_m} - b_k \tilde{e}^{-}_{k j} {\partial \tilde{T}^{i j} \over \partial k^i}. 
\ee

Putting together \eqref{polder1},\eqref{polder2}, \eqref{polderexp} in \eqref{confontwithpol} (and using the fact that $k^i e_{i j}^{-} = 0$) we find that
\be
\label{confonTsemifinal}
\begin{split}
b^{k} \tilde{K}_k \left(e^-_{i j} {T^{i j} \over \norm{k}^p}\right) &= 2 \Big[{-(p+2) \dotp[b,k] e^{-}_{i j} T^{i j} + (p+2) b^k e^{-}_{k j} k_i T^{i j} \over \norm{k}^{p+2}}+
  b_m {{e}^{-}_{i j} \over \norm{k}^p} {\partial {T}^{i j}  \over \partial k_{m}} \\ &+ b_i {{e}^{-}_{m j} \over \norm{k}^p} {\partial  {T}^{i j}  \over \partial k_m} - b^k {{e}^{-}_{k j} \over \norm{k}^p} {\partial {T}^{i j} \over \partial k^i}  \Big] + (i \leftrightarrow j) - \tilde{e}^{-}_{i j} b^k \tilde{K}_k \tilde{T}^{i j}.
\end{split}
\ee
The factor of $2$ works out by realizing that \eqref{polder1} and \eqref{polder2} give the same contribution,
and that the factor of ${1 \over 2}$ in \eqref{spindertomom} cancels
with the factor of $2$ in the definition of $\tilde{K}$.

Now, we are almost done. We just need to convert the action of $k$-derivatives on $\tilde{T}$ to the action of these derivatives on $T$. We see that
\be
\begin{split}
{\partial \over \partial k_{m}} \tilde{T}^{i j} &= {-(p+2) \over \norm{k}^{p+4}} k^{m} T^{i j} +{1 \over \norm{k}^{p+2}}  {{\partial T^{i j} \over \partial k_{m}}}, \\
 {\partial \over \partial k_n} {\partial \over \partial k_m} \tilde{T}^{i j} &=(p+2) \left[ {(p+4) \over \norm{k}^{p+6}} k^n k^m T^{i j} - { \eta^{m n} \over \norm{k}^{p+4}} T^{i j} -  {k^m \over \norm{k}^{p+4}} {\partial T^{i j} \over \partial k_n} -  {k^n \over \norm{k}^{p+4}}  {{\partial T^{i j} \over \partial k_{m}}}\right] \\ &+ {1 \over \norm{k}^{p+2}} {\partial^2 T^{i j} \over \partial k_n \partial k_m},\\
2 k_n {\partial \over \partial k_n} {\partial \over \partial k_m} \tilde{T}^{i j} &= 2(p+2)\left[{(p+3) k^m  \over \norm{k}^{p+4}}  T^{i j} 
  -  {k^m k_n \over \norm{k}^{p+4}} {\partial T^{i j} \over \partial k_n} -  {1 \over \norm{k}^{p+2}}  {{\partial T^{i j} \over \partial k_{m}}} \right] \\ &+ {2 k_n \over \norm{k}^{p+2}} {\partial^2 T^{i j} \over \partial k_n \partial k_m}\\
- k_m {\partial \over \partial k^n} {\partial \over \partial k_n} \tilde{T}^{i j} &= -(p+2)\left[{(p+4-d) k_m \over \norm{k}^{p+4}} T^{i j} 
  - {2 k_m k_n \over \norm{k}^{p+4}} {\partial T^{i j} \over \partial k_n} 
\right] - {k_m \over \norm{k}^{p+2}} {\partial^2 T^{i j} \over \partial k_n \partial k^n}.
\end{split}
\ee
All of these lead to
\be
\label{conftildeonT}
\tilde{K}_m \tilde{T}^{i j} = (p+2)(p+d) {k_m \over \norm{k}^{p+2}} T^{i j} - {2(p+1) \over \norm{k}^{p+2}} {\partial T^{i j} \over \partial k_m} + {1 \over \norm{k}^{p+2}} \tilde{K}_m T^{i j}.
\ee
Finally, putting together \eqref{confonTsemifinal} and \eqref{conftildeonT}, and using $d = 3$, we see that
\be
\begin{split}
&b^{k} \tilde{K}_k \left(e^-_{i j} {T^{i j} \over \norm{k}^p}\right) = (p+2)(p+d - 4) \dotp[b,k] {e^{-}_{i j} T^{i j} \over \norm{k}^{p+2}} \\
&  e^{-}_{i j} \left[{b_k \over \norm{k}^p} \tilde{K}_k T^{i j}  -2 p {b_k \over \norm{k}^p}  {\partial T^{i j} \over \partial k_k} + {2 \over \norm{k}^p} \left(b_m {\partial T^{m j} \over \partial k_i} - b^i {\partial T^{m j} \over \partial k^m} \right) + (i \leftrightarrow j) \right] \\ 
&+  \left({6 b^k e^{-}_{k j} k_i T^{i j} \over \norm{k}^3}\right) + (i \leftrightarrow j).
\end{split}
\ee
Now, we see that for $p = 1$, we have
\be
\label{confonTfinal}
2 b_{k} \sigma^k_{\alpha \dot{\alpha}} {\partial \over \partial \la_{\alpha}} {\partial \over \partial \lb_{\dot{\alpha}}} \left(e^-_{i j} {T^{i j} \over \norm{k}}\right) = 
 -b^k e^{-}_{i j} K_k T^{i j} +  \left({6 b^k e^{-}_{k j} k_i T^{i j} \over \norm{k}^3}\right) + (i \leftrightarrow j).
\ee
This matches precisely with Eqn. (4.37) of \cite{Maldacena:2011nz}, up to the same overall minus sign that appeared above.

The result \eqref{confonTfinal} is useful in the following way. Consider a correlation function with some number of $T$'s contracted with polarization tensors. Now, the action of the conformal generator on this object is not very well defined because the polarization tensor is {\em not}  a well defined function of the momenta. (This is because, given a polarization tensor, we can multiply it by a phase and obtain an equally good tensor.)  However, we do know that the conformal generator acting on the {\em bare correlator} (without any polarization tensor) vanishes by conformal invariance. What \eqref{confonTfinal} tells us is that if we act with the double-derivative operator (which is well-defined on polarization tensors also, as opposed to the original conformal generator) then this is the same as the action of the original conformal generator on the bare correlator (which vanishes) plus a term that is proportional to the Ward identities.

\subsection{Relation between $\tilde{R}$ and $S$ \label{tildeRS}}
Finally, let us show how spinor identities can be used to derive a
relation between $\tilde{R}$ and $S$. 
Contracting \eqref{formcorr} 
with the polarization tensor in \eqref{gravpolten} and using the fact
that $e^{\pm}_{i j}k_2^i = - e^{\pm}_{i j} k_1^i$, which
follows $e^{\pm}_{i j} k_3^i = -e^{\pm}_{i j}(k_1^i +
k_2^i) = 0$, we see that
 \be
\begin{split}
&{1 \over \norm{k_1} \norm{k_2} \norm{k_3} }e^{\pm,i j} \langle O(\vect{k}_1)
O(\vect{k}_2) T_{ij}(\vect{k}_3)\rangle \\
&= -{1 \over \norm{k_1} \norm{k_2} \norm{k_3}} e^{\pm,i j} k_{1 i} k_{2 j} \left(f_1(\norm{k_1},
  \norm{k_2}, \norm{k_3}) + f_1(\norm{k_2},
  \norm{k_1}, \norm{k_3}) - 2 f_2(\norm{k_1},
  \norm{k_2}, \norm{k_3}) \right) \\ &= -{2 \over \norm{k_1}
  \norm{k_2} \norm{k_3} } e^{\pm,i j} k_{1 i} k_{1
  j} S(\norm{k_1}, \norm{k_2}, \norm{k_3}). 
\end{split}
\ee

Now, specializing the the negative helicity polarization tensor we can write:
\be
e^{-, i j} k_{1 i} k_{2 j} = {1 \over 4 \norm{k_3}^2} \dotl[\la_3, \la_1]
 \dotlm[\la_3, \lb_1] \dotl[\la_3, \la_2] \dotlm[\la_3, \lb_2].
\ee

We can use some spinor identities to rewrite the amplitude above. These identities simply come from the conservation of momentum,
which in the spinor basis, can be written:
\be
\la_{1 \alpha} \lb_{1 \dot{\alpha}} + \la_{2 \alpha} \lb_{2 \dot{\alpha}} +  \la_{3 \alpha} \lb_{3 \dot{\alpha}} =  (\norm{k_1} + \norm{k_2} + \norm{k_3}) \sigma^0_{\alpha \dot{\alpha}}.
\ee
Contracting this with $\la_3^{\alpha} \lb_1^{\dot{\alpha}}$, this leads to:
\be
\dotl[\la_3, \la_2] \dotlb[\lb_2, \lb_1] = -  (\norm{k_1} + \norm{k_2} + \norm{k_3}) \dotlm[\la_3, \lb_1],
\ee
and we can derive a similar identity
\be
\dotl[\la_3, \la_1] \dotlb[\lb_1, \lb_2] = -  (\norm{k_1} + \norm{k_2} + \norm{k_3}) \dotlm[\la_3, \lb_2].
\ee
Moreover, we also have the identity
\begin{equation}
\begin{split}
\dotl[\la_1, \la_2] \dotlb[\lb_1,\lb_2]&= -2 \big((\vect{k_1} \cdot \vect{k_2}) -  \norm{k_1}\norm{k_2}\big) = (\norm{k_1} + \norm{k_2})^2  -  (\vect{k_1} + \vect{k_2})^2 \\ &= (\norm{k_1} + \norm{k_2} + \norm{k_3})(\norm{k_1} + \norm{k_2} - \norm{k_3}).
\end{split}
\end{equation}
Putting these relations together we immediately get \eqref{reltilders}.

\bibliographystyle{jhepmod}
\bibliography{references}

\end{document}